\documentclass[10pt,journal,compsoc]{IEEEtran}
\usepackage{amsmath}

\usepackage{color,soul}

\usepackage[table]{xcolor}
\definecolor{light-gray}{gray}{0.95}
\definecolor{dark-gray}{gray}{0.3}
\usepackage{multirow}
\makeatletter
\newcommand*\bigcdot{\mathpalette\bigcdot@{.5}}
\newcommand*\bigcdot@[2]{\mathbin{\vcenter{\hbox{\scalebox{#2}{$\m@th#1\bullet$}}}}}
\makeatother
\usepackage{comment}

\ifCLASSOPTIONcompsoc
  \usepackage[nocompress]{cite}
\else
  \usepackage{cite}
\fi
\ifCLASSINFOpdf
  \usepackage[pdftex]{graphicx}
  \graphicspath{{./figures/}}
  \DeclareGraphicsExtensions{.pdf,.jpeg,.png}
\else
  \usepackage[dvips]{graphicx}
\fi
\usepackage{amsmath}
\DeclareMathOperator\erf{erf} % Needed for Gaussian CDF formula

\begin{document}
\title{Risk of Cascading Blackouts Given Correlated Component Outages}

\author{Laurence~A.~Clarfeld,~\IEEEmembership{}
        Paul~D.H.~Hines,~\IEEEmembership{Senior~Member,~IEEE}
        Eric~M.~Hernandez,~\IEEEmembership{}
        and~Margaret~J.~Eppstein.~\IEEEmembership{}% <-this % stops a space
\IEEEcompsocitemizethanks{%
\IEEEcompsocthanksitem This work was supported in part by NSF Award Nos.~CNS-1735513, ECCS-1254549, and DGE-1144388
\IEEEcompsocthanksitem L.A. Clarfeld and M.J. Eppstein are with the Department
of Computer Science, University of Vermont, Burlington, VT, 05405 (e-mail: Laurence.Clarfeld@uvm.edu; Maggie.Eppstein@uvm.edu).
\IEEEcompsocthanksitem P.D.H. Hines is with the Department of Electrical and Biomedical Engineering, University of Vermont, Burlington, VT, 05405 (e-mail: paul.hines@uvm.edu).
\IEEEcompsocthanksitem E.M. Hernandez is with the Department of Civil and Environmental Engineering, University of Vermont, Burlington, VT, 05405 (e-mail: eric.hernandez@uvm.edu).}
}

\IEEEtitleabstractindextext{%
\begin{abstract}
Cascading blackouts typically occur when nearly simultaneous outages occur in $k$ out of $N$ components in a power system, triggering subsequent failures that propagate through the network and cause significant load shedding. While large cascades are rare, their impact can be catastrophic, so quantifying their risk is important for grid planning and operation. A common assumption in previous approaches to quantifying such risk is that the $k$ initiating component outages are statistically independent events. However, when triggered by a common exogenous cause, initiating outages may actually be correlated. Here, copula analysis is used to quantify the impact of correlation of initiating outages on the risk of cascading failure. The method is demonstrated on two test cases; a 2383-bus model of the Polish grid under varying load conditions and a synthetic 10,000-bus model based on the geography of the Western US. The large size of the Western US test case required development of new approaches for bounding an estimate of the total number of $N-3$ blackout-causing contingencies. The results suggest that both risk of cascading failure, and the relative contribution of higher order contingencies, increase as a function of spatial correlation in component failures.% Note: this is now 195 words. 200 words is the maximum suggested length for the target journal. Depending on where you look, the actual limit may be 250 words, but there may be conflicting info out there.
\end{abstract}

% OLD ABSTRACT
%\maggie{The flow of this abstract feels choppy -- need to work on the logical flow between sentences -- abstract for PSCC was better.} Quantifying the risk to electric transmission networks due to N-k contingencies that result in cascading power failure (a.k.a. ``malignancies'') is an important but challenging task, due in part to the combinatorial search space associated with k $>$ 1. \maggie{should we first define N and k?} Cascades are often triggered by ``common cause'' events, such as extreme weather, that result in multiple near-simultaneous outages of spatially proximal components. However, prior approaches have assumed uncorrelated outage probabilities. In this work, we use copula analysis, applied to malignancy sampling with ``Random chemistry'', as the bases for assessing risk due to correlated outages. We compare the risk associated with N-2 and N-3 contingencies under varying levels of spatial correlation on the 2838-bus Polish and 10,000-bus Western US test cases. We show that, while the relative risk of N-3 malignancies \maggie{malignancies may not be well-known} to N-2 malignancies increases with increasing spatial correlation, the vast majority of the risk remains attributable to N-2 malignancies at realistic levels of correlation.

\begin{IEEEkeywords}
blackout risk, cascading failure, cascading outage, correlated outages, Random Chemistry.
\end{IEEEkeywords}}

\maketitle

% is not selected <OR> if conference mode is selected - because compsoc
% conference papers position the abstract like regular (non-compsoc)
\IEEEdisplaynontitleabstractindextext

% \fi
\IEEEpeerreviewmaketitle

\ifCLASSOPTIONcompsoc
\IEEEraisesectionheading{\section{Introduction}\label{sec:introduction}}
\else
\section{Introduction}
\label{sec:introduction}
\fi

\IEEEPARstart{C}{ascading} power failures are typically initiated when a small number of $k$ components in a power system of $N$ components disconnect %stop working
nearly simultaneously, and the subsequent rerouting of power flow triggers additional component outages. This process continues until the system reaches a state of equilibrium. While most cascades do not propagate extensively throughout the network, the rare cases when they do can cause massive blackouts affecting millions of people. Due to their vast size and substantial social and economic costs, the risk they pose to power systems is significant~\cite{chen2006probability, dobson2007complex,Newman2011}.
\par

    Networks with heterogeneous load profiles, such as power systems, are particularly prone to cascades; without the right precautions, even a single node may trigger a cascade~\cite{motter2002cascade}. To mitigate the risk posed by cascading failure, power systems are required to operate such that no single component outage will cause a cascade (so-called $N$-1 security). While grid planners and operators are now also obligated to consider the risk of cascading failure due to multiple contingencies  $(k>1)$~\cite{nerc2007top}, it is not yet clear how to estimate this risk. For brevity, minimal $N-k$ contingencies that result in a cascading blackout are referred to as ``malignancies'', while contingencies that do not cause a blackout are referred to as ``benign''~\cite{eppstein2012random}. By ``minimal'', we mean that no smaller subset of outages results in a blackout. Analysis of high order malignancies is challenging due to the nonlinear ways in which cascades propagate, the vast number of $N-k$ malignancies, and the combinatorial search space of possible contingencies.

    In addition to helping to quantify risk of cascading failure, studying $N-k$ malignancies may potentially inform mitigating actions. For example, prior research into a simple model of cascading overloads in communication networks~\cite{motter2004cascade} suggests that the intentional removal of key components directly after initiating sets of outages may reduce the size of subsequent cascades. In a power system model of the Polish grid, optimally dispatching generation assuming a 50\% reduction in line limits on the 3 branches that contribute the most to the risk of cascades from $N-2$ malignancies dramatically reduces the overall risk of cascading failure with only a modest increase in operational costs~\cite{rezaei2015rapid}.\par

Many previous approaches to cascading failure risk analysis (including our own) assumed initiating component outages to be independent events~\cite{chen2006probability, rezaei2015rapid, rezaei2015estimating, kock2014probabilistic, rezaei2014estimating}. However, $N-k$ malignancies triggered by the same exogenous event, or ``common cause'', represent a significant source of risk to power systems~\cite{papic2017research}, and can result in spatial correlation in initiating outages. For example, extreme weather events can result in spatially correlated damage~\cite{dobson2017exploring}, protection system failures can sometimes cause multiple outages within a small geographic region~\cite{jiang2011new}, and terrorist attacks may be spatially localized, such as in the 2013 sniper attack on the Metcalf substation near San Jose, CA, where the perpetrators shot 17 transformers at the same substation~\cite{smith2014us}. Non-spatial attributes, such as component age, may also induce correlations in component failures~\cite{li2002incorporating, salman2014age}.

There is a dramatically increasing computational burden to assessing risk for $N-k$ malignancies as $k$ increases, so it is important to understand the degree to which higher-order ($k>2$) malignancies contribute to risk, and thus how important it is to consider them in risk estimation. Even though there are many more $N-3$ than $N-2$ malignancies for any given system, when there is no correlation in initiating outages the probability of $N-3$ malignancies occurring is so much lower than that of $N-2$ malignancies that the impact of $N-3$ malignancies on risk is negligible~\cite{rezaei2015estimating}. However, as correlation in component outages increases, the impact of higher order malignancies on risk will increase. To what degree should risk analysis take into account the conditional probability of component failure, given a common cause?

There has been some %amount of
prior work on ways to incorporate correlation into risk analysis. In~\cite{bernstein2014power}, correlation was incorporated by assuming 100\% correlation of outages within a fixed radius. In~\cite{dobson2017exploring}, spatial correlation was achieved by determining outage rates of lines adjacent to initial failures probabilistically, according to a Poisson process. In~\cite{scherb2017reliability}, a random field with spatial autocorrelation was used in a cascade model to assess risk from common-cause events. Others~\cite{chen2005cascading} have simulated the impact of hidden relay failures on cascading failure risk by allowing proximate lines to trip probabilistically. \par

Another approach to incorporating correlation into risk estimation is {\em via} copula analysis. Popularized in the field of finance~\cite{cherubini2004copula}, copulas have been used in a wide variety of disciplines to model the co-dependence of multiple variables~\cite{onken2009analyzing,schoelzel2008multivariate}.  Within the realm of power systems, copulas are a popular tool for uncertainty analysis. They have been used in the modelling of stochastic generation, such as wind~\cite{hagspiel2012copula,papaefthymiou2008modeling,papaefthymiou2009using}. The impacts of variable infeeds on security assessment have also been considered using copulas~\cite{de2018framework}. Li~\cite{li2014risk} suggests copulas as a useful way to incorporate correlation between random variables in power systems risk analysis. \par

% However, to our knowledge, copula analysis has not yet been applied to cascading failure risk analysis, outside of the preliminary study~\cite{clarfeld2018assessing} that we build upon here.\par

A flexible and generalizable approach to risk estimation given correlated component outages was presented in~\cite{clarfeld2018assessing} and used to estimate risk due to $N-2$ malignancies in a 2383-bus model of the Polish grid. This paper extends that work in several significant ways including: (i) incorporating the effects of $N-3$ malignancies, (ii) studying how the risk due to $N-3$, relative to $N-2$, malignancies changes as a function of correlation in outage probabilities, and (iii) applying the method to a much larger and more geographically realistic 10,000-bus test case, which necessitated (iv) development of new methods for estimating the total number of $N-3$ malignancies.

This paper is organized as follows: methods for risk estimation using samples of $N-k$ malignancies, and the computationally efficient ``Random Chemistry'' (RC) sampling method used in this work, are reviewed in Sections~\ref{sec:estimating_risk} and~\ref{sec:RC}, respectively. In Section~\ref{copula} a method using copula analysis to incorporate initiating outage correlations into risk estimation is presented, and in Section~\ref{sec:distance} an approach to quantifying distance between transmission lines, when considering spatial correlation, is described. The two test cases used to demonstrate the method are described in Section~\ref{sec:casestudies}; new methods for bounding the total number of $N-3$ malignancies in large systems are described in Section~\ref{sec:num_of_maligs}. Results and discussion are presented in Sections~\ref{sec:results} and~\ref{sec:discussion}, respectively.

% \hfill lac
% \hfill October 1, 2018

%\IEEEpubidadjcol

\section{Methods}
\subsection{Estimating Risk of Cascading Failure} \label{sec:estimating_risk}

This study uses the method for estimating risk of cascading failure from sampled $N-k$ malignancies presented in~\cite{rezaei2014estimating,rezaei2015estimating}, briefly reviewed below.

The risk due to a set of branches (transmission lines or transformers) $\omega$ can be calculated as ~\cite{vaiman2012risk}:
\begin{equation}
R_{\omega } = p_{\omega }s_{\omega }
\end{equation}\label{Romega}%
where $p_{\omega }$ is the joint probability of the branches in $\omega$ failing and $s_{\omega }$ is the size of the resultant blackout. Note that $p_{\omega}$ is itself a function of $p_i$, the independent outage probability for each branch $i \in \omega$, as well as any effect of correlation among branch outage probabilities (as further defined in Section~\ref{copula}).

Blackout size $s_{\omega }$ is quantified as the total power (MW) unserved due to load shedding. In this work, a cascading blackout is considered to have occurred when 5\% or more of the total load is shed in DCSIMSEP, a simulator of cascading outages in power systems~\cite{Hines2016cascadingChapter}. The risk posed to the system by all $N-k$ malignancies comprising branches $\omega$, for a given $k$, is then:
\begin{equation}\label{Rk}
R_{k} = \sum_{\omega  \in \Omega_k }R_{\omega}
\end{equation}
where $\Omega_k$ is the complete set of all $N-k$ malignancies for the specified $k$. For realistically-sized power systems it is not computationally tractable to find the entire set $\Omega_k$ for $k>2$. However, if $\Omega_k^{sampled} \subset \Omega_k$ is a large and representative subset of size $|\Omega_k^{sampled}|$, comprising all unique $N-k$ malignancies found by many iterations of some sampling strategy, and if the size of the complete set of $N-k$ malignancies $|\Omega_k|$ can be estimated, then risk $\hat{R_k}$ associated with $N-k$ malignancies, for a given $k$, can be estimated as follows:
\begin{align}\label{Rk_est}
    \hat{R_k} = \frac{|\Omega_k|}{|\Omega_k^{sampled}|} \sum_{\omega \in \Omega_k^{sampled}} R_{\omega}
\end{align}
Estimating $|\Omega_k|$ for $k>2$ on large systems is itself a very challenging problem, as discussed in Section~\ref{sec:num_of_maligs}.

Considering only malignancies with $k \le k_{max}$ and assuming that non-minimal supersets of malignancies do not substantially change the amount of load shed, as justified in ~\cite{rezaei2015estimating}, the risk of cascading failure can be approximated as:
\begin{equation} \label{risk_eq}
\hat{R} = \sum_{k  \in \{2..k_{max}\}}\hat{R_k}
\end{equation}
In this work, $k_{max}=3$.

\subsection{Random Chemistry Sampling} \label{sec:RC}

For each $k$, there are ${N \choose k}$ possible $N-k$ contingencies, only a small proportion of which are malignancies. While exhaustive search may be feasible (albeit time consuming) for $k=2$, it is computationally intractable for $k>2$ in large power systems. Thus, many iterations of the Random Chemistry (RC) sampling method were used to efficiently identify large sets of $N-k$ malignancies in each test case. RC is a stochastic set size reduction algorithm for identifying a small minimal set of initiating events that trigger some outcome of interest~\cite{eppstein2007genomic}, and was first applied for identifying $N-k$ malignancies in power systems in~\cite{eppstein2012random}. For the reader's convenience, the RC algorithm is briefly reviewed below.

The RC algorithm uses a subset reduction scheme $\{a_1,a_2,\dots,a_{final}\}$. Subsets of size $a_1$ are randomly sampled from a universal set of $N$ system components until one such set is found that causes a blackout; if $a_1$ is relatively large, this typically requires few tries. A set of size $a_{i+1}$ is then randomly sub-sampled from the preceding set of size $a_i$ until a set is found that causes a blackout (or some maximum number of sub-samples is tried, in which case the algorithm aborts), and so on for each subsequent set size in the scheme.

If $a_{i+1} = a_i/c$, for some constant $c$, then the algorithm requires only $O(N \log N)$ time to identify a subset of size $a_{final}$. As in~\cite{eppstein2012random}, we use $c=2$ from $a_1$ down to subsets of size 20 and then use $c=1.5$ down to $a_{final}$. A  bottom-up brute force search of all subsets of a given size $k$ is subsequently applied (conducted in randomized order, starting from $k=2$), exiting when the first minimal malignancy of size $k=\{2,3, \dots, a_{final}\}$ is identified.

% Suggestion for end of the above paragraph:
% ...and then use $c=1.5$ down to $a_{final}$. A bottom-up brute force search of all subsets of $a_{final}$ is then conducted (in randomized order, starting from $k=2$), exiting when the first minimal malignancy of size $k=\{2, \dots k_{max}\}$ is identified.

Repeated sampling with independent RC trials is performed to compile large subsets of $N-k$ malignancies ($\Omega^{RC}_k$). Risk due to each $k \le k_{max}$ is then calculated using $\Omega^{sampled}_k = \Omega^{RC}_k$ in (\ref{Rk_est}) for estimating system risk with (\ref{risk_eq}).

%(Fig.~\ref{fig:RC}). \\

%Random Chemistry figure:
%\begin{figure}[!hb]
%\includegraphics[width=3.5in]{RC_algorithm_schematic}
%\caption{Schematic of the RC algorithm applied to finding a minimal $N-k$ malignancy in a power system (abstracted from~\cite{eppstein2012random})\maggie{replace $2\rightarrow 5$ with $2\rightarrow 3$ and show an arrow leaving the last box to $\{N-2,N-3\}$}.
%The specific values of the initial ($80$) and final ($k\leq5$) contingency set sizes are user-specified parameters.
%}\label{fig:RC}
%\end{figure}

% In ~\cite{eppstein2012random} we applied RC to the problem of identifying minimal $N-k$ malignancies that lead to cascading failures in power grids (Fig.~\ref{fig:RC}), where $N$ is the total number of branches in the power system.

%Using the approach described above, we previously showed that RC can be used to efficiently estimate the system-wide risk of large cascading failures in power grids ~\cite{rezaei2014estimating,rezaei2015rapid,rezaei2015estimating}.
A comparison of risk estimation using RC sampling {\em vs.} Monte Carlo (MC) sampling on a model of the Polish power system at peak winter load~\cite{zimmerman2011matpower} showed that the RC approach was at least two orders of magnitude faster than MC on this system, and did not introduce measurable bias into the estimate~\cite{rezaei2014estimating,rezaei2015estimating}. However, these previous studies assumed branch outages were uncorrelated. Under that assumption, $N-3$ malignancies contribute relatively little to the risk, despite the fact that $|\Omega_3^{RC}| \gg |\Omega_2^{RC}|$, since their probability of occurrence is so much smaller than that of the $N-2$ malignancies~\cite{rezaei2015estimating}. \par

In this study, the universal set is assumed to comprise the set of $N$ branches in each test case, $a_{final}=5$, and up to 20 sub-samples at each set size were allowed before aborting an RC trial, as in~\cite{rezaei2014estimating,rezaei2015estimating}. The specific set reduction scheme used for each test case is given in Section~\ref{sec:casestudies}. Note that $k_{max} < a_{final}$ because, with the number of RC trials performed, $|\Omega^{RC}_k|$ was insufficient for estimating $|\Omega_k|$ for $k>3$.

% This commented paragraph is replaced by section~\ref{sec:num_of_maligs}
% The parametric curve-fitting method proposed in~\cite{rezaei2014estimating} was found to be ineffective in deriving estimates for $\Omega_k$ for our new test case. Here, we leverage the fact that this is the analogous problem to population size estimation in conservation biology. Literature from this field suggests that non-parametric methods would be more effective at finding $\Omega_k$, but can only achieve a lower-bound estimate~\cite{chao2014species}. Through experimentation, we found that a Abundance-based Coverage Estimator approach (ACE-1) to yield the best estimates for simulations of sampling methods with similar bias to Random Chemistry~\cite{chao1992estimating}.\par

\subsection{Copula Analysis for Correlation}\label{copula}

%In~\cite{clarfeld2018assessing} a method was introduced using copula analysis to generalize the risk estimation approach to systems with correlated outages, and applied it to the 2383-bus Polish system, with $k_{max}=2$. Here, this method is extended to $k_{max}=3$ and a study was conducted to asses how the relative risk of $N-3$ malignancies changes as a function of correlation in outage probabilities, both in the Polish system and a larger 10,000-bus test case, described in Section~\ref{sec:casestudies}. \par

Copula functions couple multivariate distributions to the marginal distributions of individual variables~\cite{nelsen_introduction_2010}. Given a set of $k$ random variables, $\mathbf{X} =[X_1,X_2,\dots,X_k]$, where $\Pr(X_i \leq t_i)$ is the marginal probability that branch $i$ fails, for some threshold $t_i$, then
% \begin{align}
% F_X(\mathbf{T}) = \Pr\left(\bigcap_{i=1}^k X_i \leq t_i\right)
% \end{align}
\begin{align}\label{eq:CDF1}
F_X(\mathbf{T}) = \Pr\left(\bigcap_{i=1}^k X_i \leq t_i\right)
\end{align}
where $\mathbf{T} = \left[t_1,t_2,\cdots,t_k\right]$, represents the joint probability that $k$ branches fail together. Without loss of generality, it is assumed that $t_i = 0$ for all $i$. \par

\begin{figure}[!ht]
\centering
\includegraphics[width=3.5in]{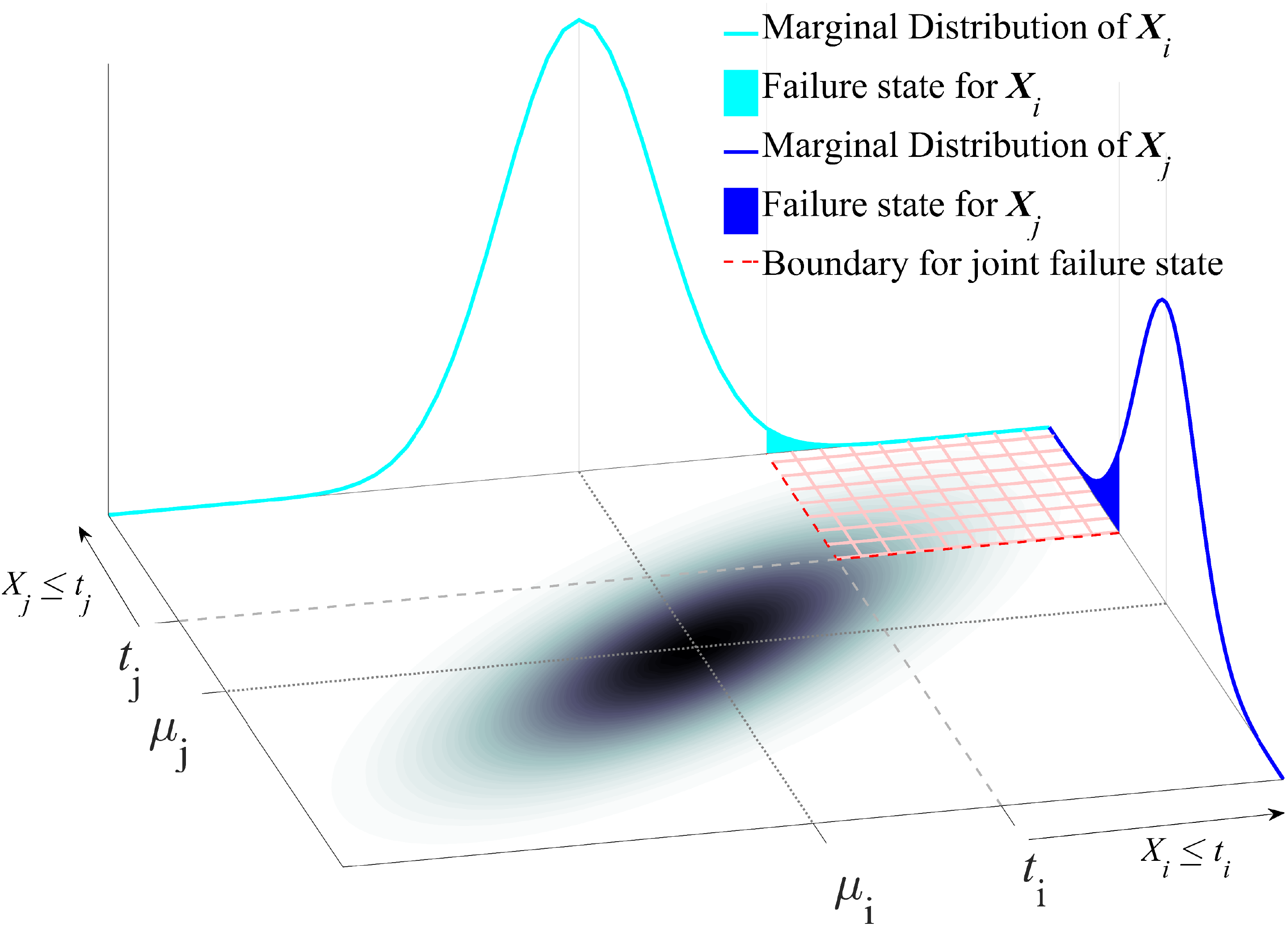}
\caption{A visual depiction of the copula method for two components $i$ and $j$  with hypothetical Gaussian distributions of some performance attributes, $X_i$ and $X_j$, which impact whether each component is operational or in a failure state. The curves on the vertical planes represent the marginal distributions of each component's attribute, with the shaded regions of these curves, ($X_i \leq t_i$) and ($X_j \leq t_j$), representing the failure state for each component. The shaded gradient on the horizontal plane represents the density of the joint distribution (copula) of the two variables, with darker shading representing higher probability density. The probability mass within the red hatched area represents the region of joint failure ($\mathbf{X} \leq [t_i, t_j]$), with the red dotted line depicting the boundaries of this region.}
\label{corr_probs_ex}
\end{figure}

There are numerous classes of copula functions in popular use. For this demonstration of the method, a Gaussian copula was assumed, but alternative distributions may be assumed where appropriate. Here, it is assumed that the inverse stress on a transmission line $i$ is a univariate Gaussian random variable $X_i=\mathcal{N}(\mu_i,\sigma_i)$, with mean $\mu_i$ and standard deviation $\sigma_i$, with the cumulative distribution function:
% no paragraph before or after equations
\begin{align}\label{eq:cdf}
F_{X_i}(x_i) = \frac{1}{2}\left[1 + \erf\left(\frac{x_i-\mu_i}{\sigma_i \sqrt{2}}\right)\right]
\end{align}
%
%Without loss of generality, we assume $\mu_i = 1$ and $x_i=0$ for all $i$.
Given the independent probability $p_i$ of branch $i$ going out, $\mu_i$ and $\sigma_i$ are chosen such that when $X_i < 0$, the branch goes out. In other words, $\mu_i$ and $\sigma_i$ are chosen such that $F_{X_i}(0)=p_i$ for each branch $i$. Without loss of generality, it is assumed that $\mu_i=1, \forall i$, and each $\sigma_i$ is then solved for as follows:
\begin{align}\label{eq:var}
\sigma_i = \frac{-1}{\erf^{-1}(2p_i-1)\sqrt{2}}
\end{align}
A multivariate normal distribution $\mathbf{X} =\mathcal{N}(\boldsymbol{\mu},\mathbf{C})$ with mean $\boldsymbol{\mu} = [\mu_1,\mu_2,\dots,\mu_k]$ and covariance matrix $\mathbf{C}$ is then used as the copula function to couple these univariate marginal distributions (Fig.~\ref{corr_probs_ex}).

\begin{figure}[!t]
\centering
\includegraphics[width=3.2in]{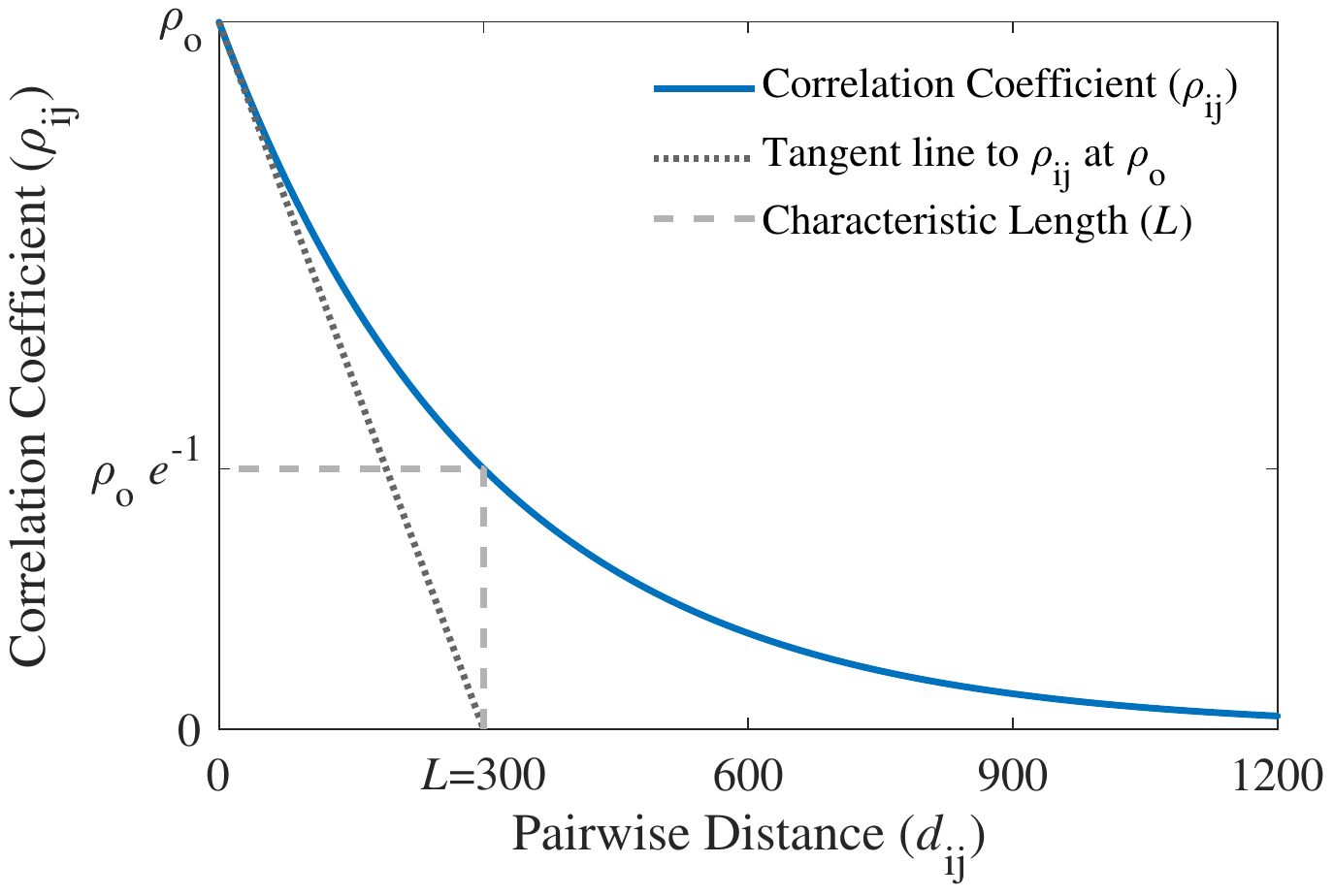}
\caption{Change in the correlation between two branches as a function of the distance between them, assuming \eqref{eq:corrcoeff} with characteristic length $L=300$ km and correlation $\rho_o$ for branches that are 0 km apart.}
\label{corr_coeff}
\end{figure}

In this study, it is assumed that the correlation between outages in branches $i$ and $j$ decays exponentially with the distance between them $d_{ij}$, according to:
\begin{align}\label{eq:corrcoeff}
\rho_{ij} = \rho_o e^{-d_{ij} / L}
\end{align}
% no paragraph
where $\rho_o$ represents the maximum possible correlation coefficient (at distance zero) and $L$ represents the characteristic length, which controls the decay rate of the correlation; $L$ can be interpreted as the distance at which $\rho_{ij}$ reaches $\rho_o/e$ (i.e., $\approx 36.8\%$ of $\rho_o$) (Fig.~\ref{corr_coeff}).

Eq.~(8) can be adjusted to represent a wide range of exogenous common cause events individually, or in combination, by adjusting the parameters $\rho_o$ and $L$ to align with data for a particular set of threats. The exponential decay form captures the spatially decaying nature of earthquakes~\cite{jayaram2009correlation}, and can approximately capture the impact of other threats that are likely to be geographically correlated, such as tornados~\cite{wurman200530} and hurricanes~\cite{willoughby2006parametric}.

The resulting correlation coefficient $\rho_{ij}$ calculated by \eqref{eq:corrcoeff}, and the standard deviations $\sigma_i$ and $\sigma_j$ calculated by \eqref{eq:var}, are used to calculate the pairwise covariance between branches $i$ and $j$ as:
\begin{align}\label{eq:cov_ij}
\mathrm{cov}(i,j) = \rho_{ij} \sigma_i \sigma_j
\end{align}
Using \eqref{eq:cov_ij} to find each element of the covariance matrix $\mathbf{C}$, the probability density function of the multivariate normal distribution \eqref{eq:mv_pdf} is used to form the copula.
\begin{align}\label{eq:mv_pdf}
f(\mathbf{x}) = \frac{1}{\sqrt{(2\pi)^{k}|\mathbf{C}|}}\exp\left\{-\frac{1}{2}(\mathbf{x}-\boldsymbol{\mu})^{\top}\mathbf{C}^{-1}(\mathbf{x}-\boldsymbol{\mu})\right\}
\end{align}
Integrating (\ref{eq:mv_pdf}) over the region in the joint distribution that represents outages of all system components gives:
\begin{align}\label{eq:jointprob}
F_X(\mathbf{0}) = \int\displaylimits_{-\infty}^{0} \int\displaylimits_{-\infty}^{0} \dots \int\displaylimits_{-\infty}^{0} f(x_1,x_2,\dots,x_k) \, \mathrm{d}x_1 \,\mathrm{d}x_2 \dots \,\mathrm{d}x_k
\end{align}
where $F_X(\mathbf{0})$ represents the joint outage probability of $k$ components $\Pr(\mathbf{X} \leq \mathbf{0})$. The multiple-integral in \eqref{eq:jointprob} represents the generalized solution for arbitrary $k$ and is equivalent to the cumulative distribution function of the multivariate normal distribution, which can be solved efficiently in MATLAB using methods described in~\cite{genz2004numerical}. In this work, $k \in \{2, 3\}$. % and solve the resultant double-integral numerically using the vectorized adaptive quadrature method~\cite{shampine_vectorized_2008}.

\subsection{Defining Inter-Branch ``Distance''}\label{sec:distance}
\begin{figure}[!ht]
\centering
\includegraphics[width=2.2in]{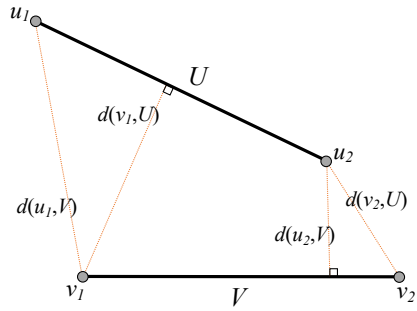}
\caption{Visual example for calculating the distance between branches $U$ and $V$ with endpoints ($u_1,u_2$) and ($v_1$,$v_2$), respectively.}
\label{br_dist_ex}
\end{figure}

The definition of ``distance'' will vary based on the type of common cause threatening the system. Assuming spatial correlation in branch outages here, without loss of generality a modified version of the inter-branch distance metric defined in~\cite{clarfeld2018assessing} is used. Branches are assumed to be straight lines between the buses that form their endpoints. Given branch $U$ with endpoints ($u_1,u_2$) and branch $V$ with endpoints ($v_1,v_2$), let the distance from $U$ to $V$ be defined as
\begin{align}\label{eq:dist}
\mathrm{Dist}(U,V) = \frac{\sum_{i=1}^2 d(u_i,V) + \sum_{i=1}^2 d(v_i,U)}{4}
\end{align}
where $d(v_i,U)$ is the minimum Euclidean distance from the point $v_i$ to the line segment $U=(u_1,u_2)$, calculated as:
\begin{align} \label{point_to_line_seg}
    d(v_i,U)=\left\{
            \begin{array}{ll}
                ||v_i - u_1|| & t \leq 0\\
                ||v_i - (u_1 + tm)|| & 0 \leq t \leq 1\\
                ||v_i-u_2|| & t \geq 1
            \end{array}
        \right.
\end{align}
where $m = u_2 - u_1$ and $t = \frac{(v_i-u_1) \, \bigcdot \, m}{||m||^2}$, as illustrated in Figure~\ref{br_dist_ex}.

\begin{figure}[!ht]
\centering
\includegraphics[width=2.5in]{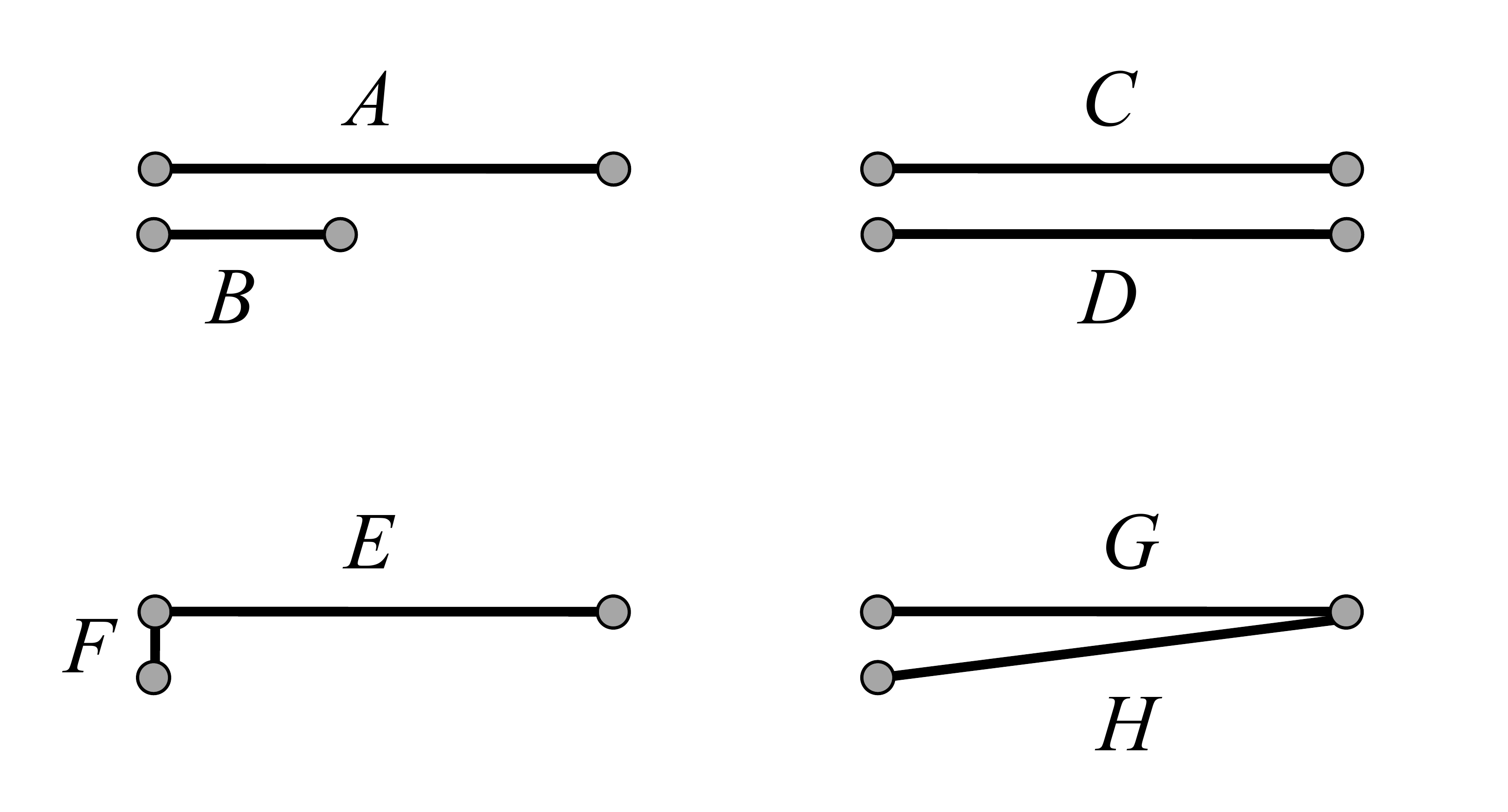}
\caption{Branch pairs used for pairwise distance examples described in the text.}
\label{br_exs}
\end{figure}

This formulation defines a semi-metric since the triangle inequality does not hold in some cases. However, all other formal requirements of a metric are met. Specifically:
\begin{enumerate}
    \item $Dist(U,V) \geq 0$
    \item $Dist(U,V) = Dist(V,U)$
    \item $Dist(U,V) = 0 \iff U = V$
\end{enumerate}
The third identity implies branches $U$ and $V$ share the same endpoints, thus are parallel. This $Dist(U,V)$ measure is consistent with what would be intuitively expected when considering spatially correlated damage. For example, in Fig.~\ref{br_exs}, ${Dist}(A,B) > {Dist}(C,D)$ and ${Dist}(E,F) > {Dist}(G,H)$.

\begin{figure}[!ht]
\centering
\includegraphics[width=2.9in]{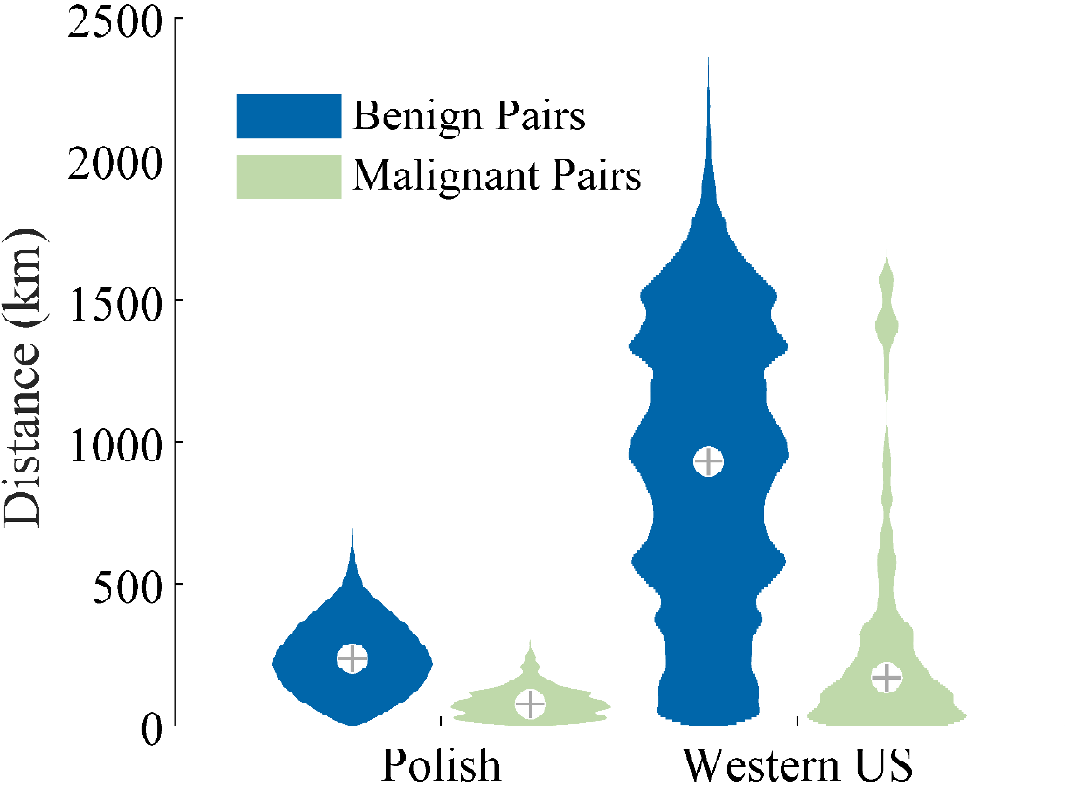}
\caption{Distance between the 540 and 564 branch pairs that form $N-2$ malignancies in the Polish and Western US test cases {\em vs.} a random sample of 1,000,000 benign branch pairs from each test case. For clarity, medians are marked with crosshairs and each distribution has been independently normalized to the same maximum width.}
\label{malig_dist}
\end{figure}

Using the $Dist(U,V)$ measure, it is apparent that branch pairs that form malignancies are much closer together than those of benign contingency pairs in both test cases (Fig.~\ref{malig_dist}). This property will exacerbate the effects of spatial correlation on risk of cascading failure.

\subsection{Case Studies}\label{sec:casestudies}
This risk estimation approach is demonstrated on two publicly available test cases, modeling the Polish and Western United States (US) transmission systems.

\begin{figure}[!t]
\centering
\includegraphics[width=2.9in, height=2.5in]{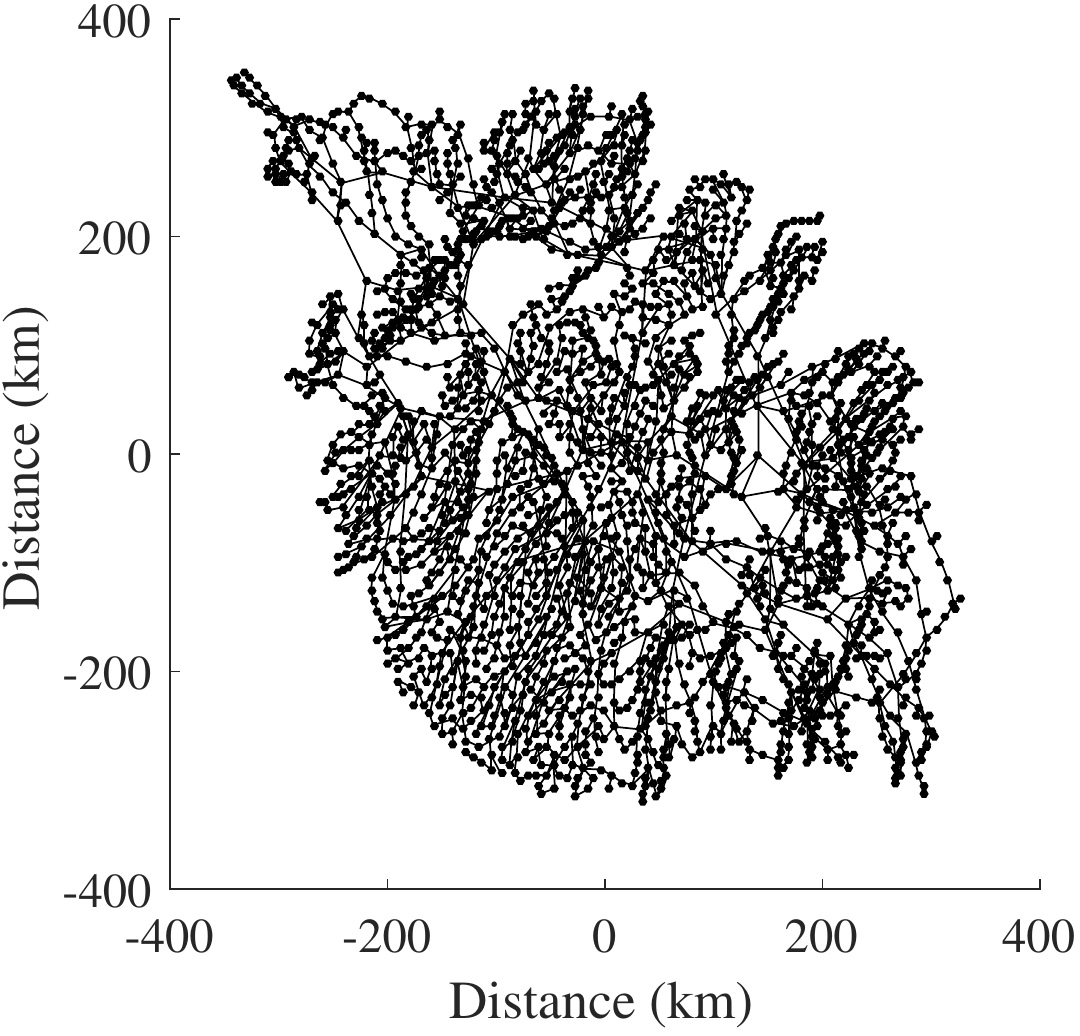}
\caption{Synthetic geographic layout of the Polish test case. Positionally, this layout is arbitrary and has been centered at (0,0), however, units were scaled so that the diameter of the geographic layout is roughly equal to that of Poland (in km).}
\label{polish_grid}
\end{figure}

The Polish test case, examined in previous work on risk estimation~\cite{rezaei2015rapid,rezaei2015estimating,clarfeld2018assessing}, contains 2383 buses and 2896 branches at a peak winter load and is distributed with the MATPOWER simulation package~\cite{zimmerman2011matpower}. The true spatial locations of branches and buses are not publicly available for this test case, so hypothetical locations were inferred based on a graph layout of the grid topology, assuming branches are straight lines between buses (Fig.~\ref{polish_grid}). This layout was then scaled to $670 \times 670$ km, the approximate width/height of Poland, to simulate geographic distances. Some of the transmission lines were overloaded in the Polish test case provided by~\cite{zimmerman2011matpower}, so the adjusted base case described in~\cite{rezaei2015rapid} was used. Unless otherwise stated, references to the ``Polish test case'' refer to this adjusted base case.  As in~\cite{rezaei2015estimating}, different load levels were modeled in the Polish test case from 80\% to 115\% of the adjusted base case by multiplying all line loads by a scalar factor and then re-running the security constrained optimal power flow, to ensure the pre-contingency system at each load level is $N-1$ secure.

\begin{figure}[!t]
\centering
\includegraphics[width=3.4in]{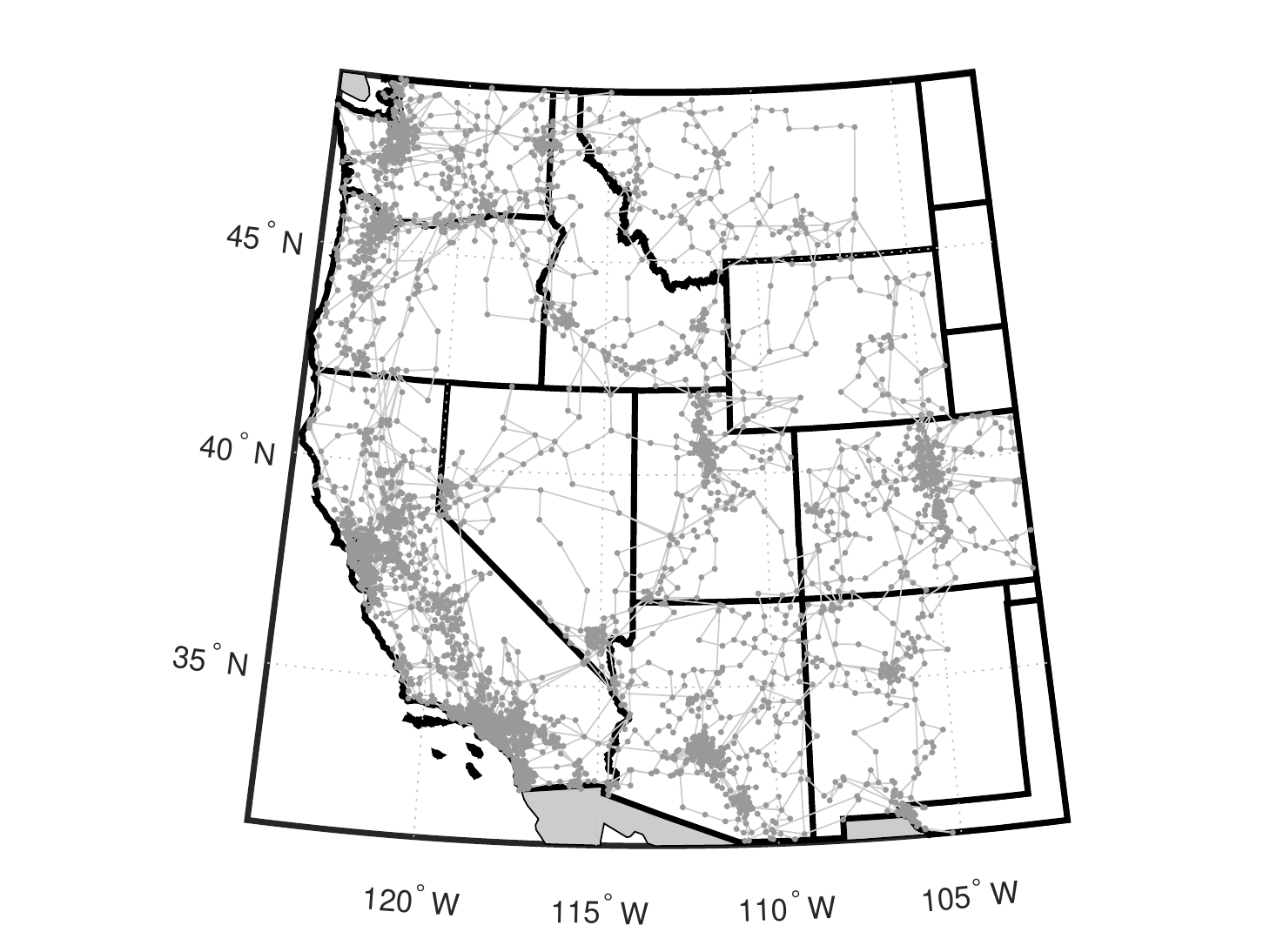}
\caption{Geographic layout of the synthetic 10,000 bus Western US test case. }
\label{western_US_grid}
\end{figure}

The Western US test case is a synthetic network based on the footprint of the western Unites States and comes {\em via} the Electric Grid Test Case Repository~\cite{birchfield2017grid}. This test case is much larger than the Polish test case, with 10,000 buses and 12,706 branches, and has a more realistic geographic layout (Fig.~\ref{western_US_grid}). As with the Polish test case, some transmission lines were overloaded for the Western US test case provided by~\cite{birchfield2017grid}, and so adjustments were made as described in~\cite{rezaei2015rapid}.
Since the case did not include short and long-term emergency flow limits (``RateB'' and ``RateC''), they were synthesized to be 110\% and 150\% of normal (``RateA'') limits, respectively.\par

Independent branch outage rates were not available for either the Polish or Western US test cases. For the results presented here, all independent outage rates were set equal to the mean outage rate of 0.9158 hours per year provided by the RTS-96 test case~\cite{force1999ieee}. These independent outage rates were deliberately assumed identical for all branches in order to more clearly elucidate the impact of spatial correlations in outage rates, as assessed using (\ref{risk_eq}), for all combinations of $L \in \{0,100,200,300\}$ km and $\rho_o \in \{0.00,0.05,0.10,0.15\}$.
% [SHOULD I ADD HERE THAT LINES WITHOUT ``B'' OR ``C'' RATES WERE SET TO 1.1x AND 1.5x ``A'' RATES? AND LINES WITH NO LIMITS WERE OK BECAUSE ELECTRICAL DISTANCE WAS $\sim$ 0?].\\

For the results shown here, the RC algorithm used the subset reduction scheme $\{80,40,20,14,10,7,5\}$ for  the Polish model, as in~\cite{eppstein2012random,rezaei2014estimating,rezaei2015rapid, rezaei2015estimating,clarfeld2018assessing}. For the larger Western US model, the initial RC subset size $a_1$ was raised to 320 to increase the probability that the initial subset causes a blackout; thus, the Western US test case used the subset reduction scheme $\{320, 160, 80, 40, 20, 14, 10, 7, 5\}$. We did 1,000,000 RC trials for the Polish test case and 704,400 RC trials for the Western US model; fewer RC trials were used for the larger test case because computation time was much higher than for the Polish model (averaging 9.5 seconds per RC trial for the Western US model {\em vs.} 2.35 seconds for the Polish model, on an Intel Core i5-3470 CPU @ 3.2GHz with 8 GB of RAM).  \par

\subsection{Estimating $|\Omega_k|$} \label{sec:num_of_maligs}

\begin{figure}[!ht]
\centering
\includegraphics[width=3.2in]{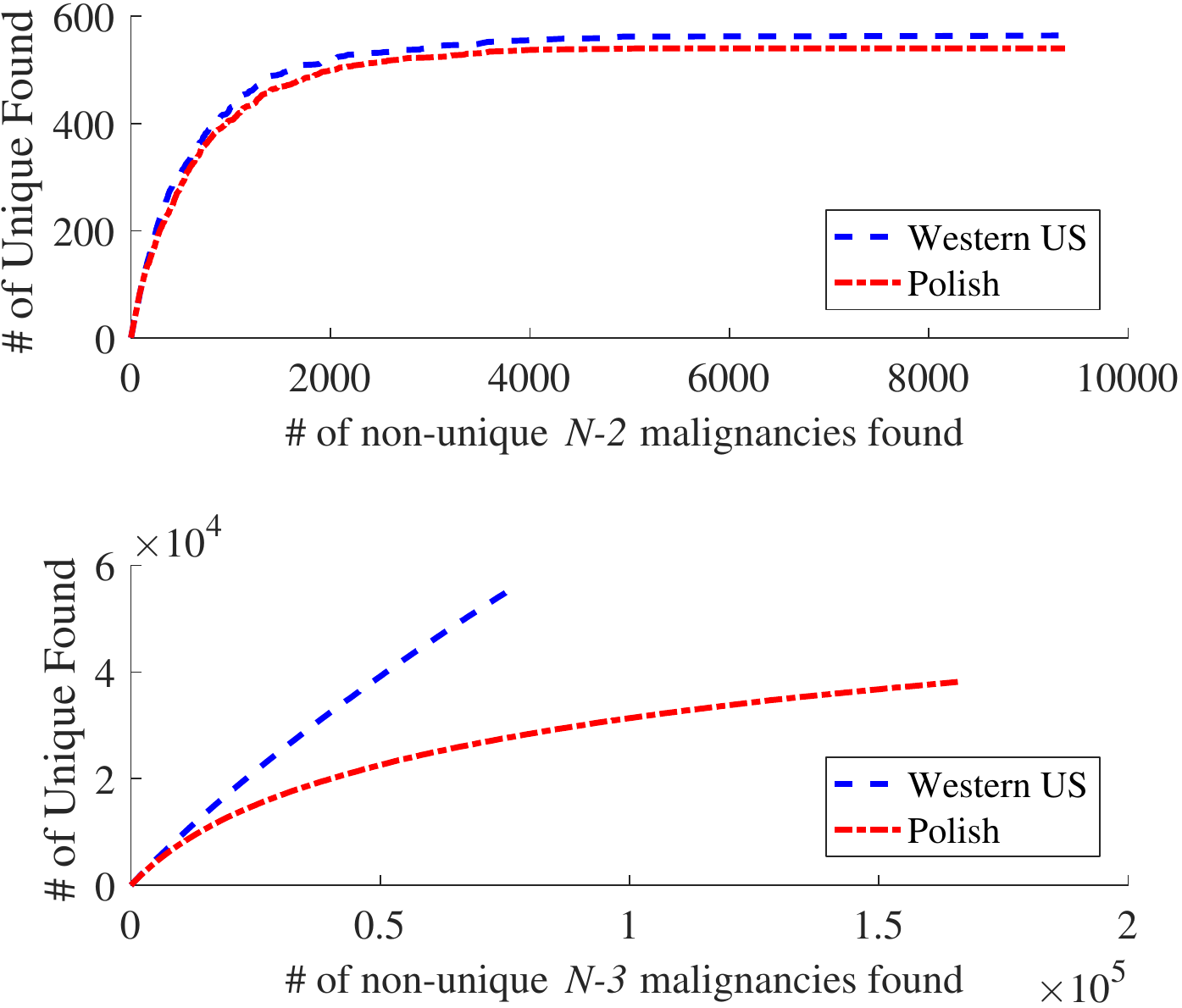}
\caption{(top) Accumulation curves for RC sampling of $N-2$ malignancies in the Polish and Western US test cases. In both cases, $|\Omega^{RC}_2| = |\Omega_2|$; (bottom) Accumulation curves for $N-3$ malignancies found by RC sampling in the Polish and Western US test cases. In both cases, $|\Omega^{RC}_3| \ll |\Omega_3|$.}
\label{accum_curves}
\end{figure}

As described in Section~\ref{sec:estimating_risk}, this approach to risk estimation requires an estimate of the total number of $N-k$ malignancies $|\Omega_k|$, for $k \le k_{max}$. There was no need to estimate $|\Omega_2|$, since RC sampling identified the complete set of $N-2$ malignancies $\Omega_2$ in both test cases, as evidenced by the flattening in the accumulation curves (Fig.~\ref{accum_curves} (top)), and later verified through brute force search for the Polish test case. The set of unique $N-2$ malignancies was complete after only 5,090 non-unique $N-2$ malignancies had been found by RC sampling in the Polish test case (of 4,191,960 possible $N-2$ contingencies) and after only 9,364 non-unique $N-2$ malignancies had been found by RC sampling in the Western US test case (of 80,714,865 possible $N-2$ contingencies).

However, obtaining the entire set of N-3 malignancies is not computationally tractable in either test case, due to the sheer size of these sets. It was initially argued (incorrectly) in~\cite{eppstein2012random} that, if one has already identified $i$ of the $N-k$ malignancies using independent RC trials, then the probability that the next identified $N-k$ malignancy has not previously been found is $(|\Omega_k|-i)/|\Omega_k|$, so one could infer $|\Omega_k|$ from the observed frequency with which unique malignancies were found (assuming sufficient curvature in the accumulation curve). However, the assumption that independent RC trials uniformly sample from the $\Omega_k$ sets has since proven false. In subsequent studies it was discovered that the accumulation curves were not exponential (as they would be if the sampling were uniform), but could be better fit with a 4-parameter exponential Weibull curve to estimate $|\Omega_k|$~\cite{rezaei2015estimating}.  While this non-linear curve-fitting approach works for estimating $|\Omega_3|$ in the Polish test case, the Western US test case is so much larger that there is insufficient curvature in the $N-3$ accumulation curve (Fig.~\ref{accum_curves} (bottom)) to reliably fit a curve.

\begin{figure}[!ht]
\centering
\includegraphics[width=3.2in]{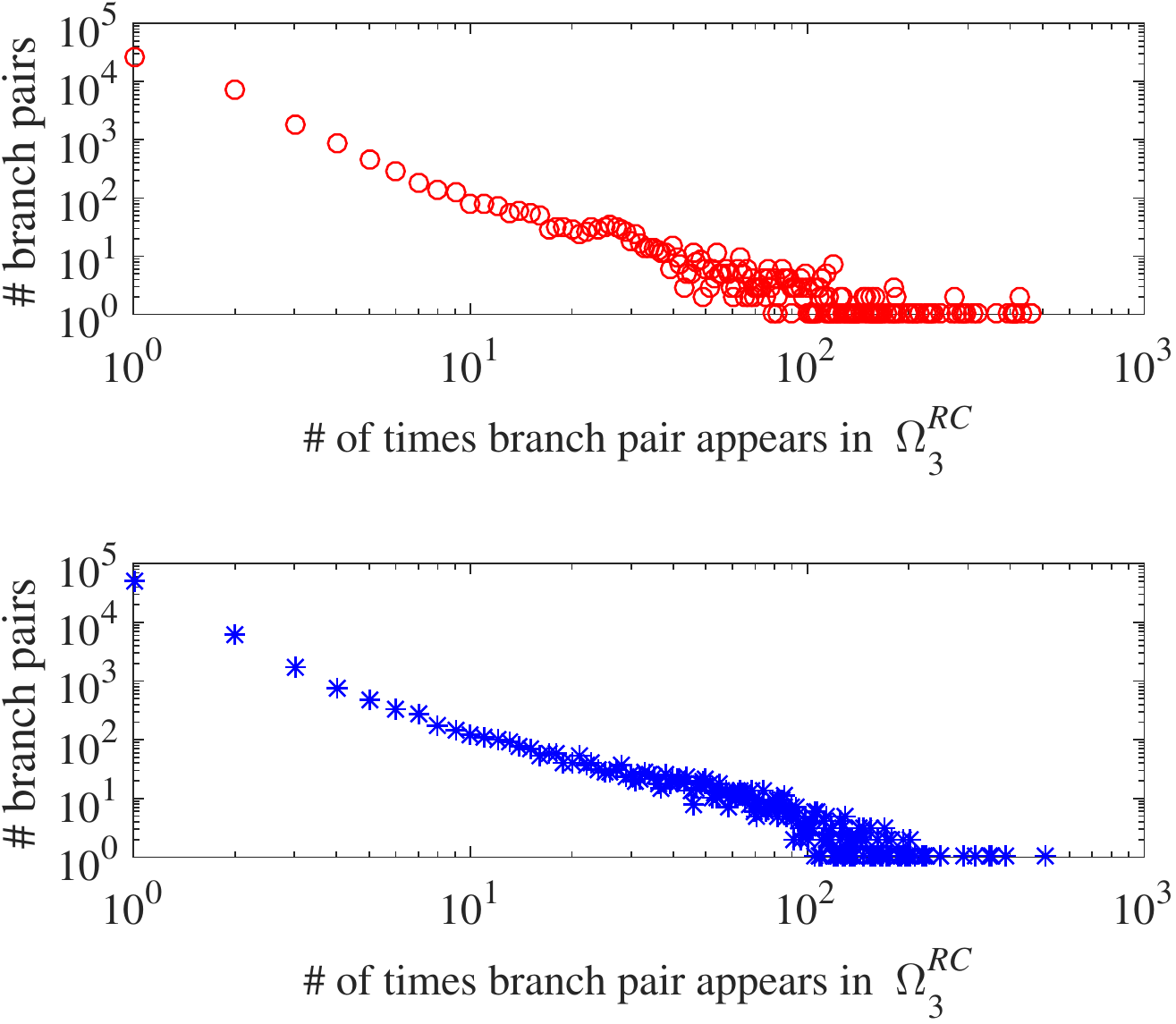}
\caption{Histograms of the number of occurrences of distinct branch pairs in unique $N-3$ malignancies found {\em via} RC in (top) the Polish test case and (bottom) the Western US test case.
}
%\maggie{This looks much better. However, it strikes me as odd that in Fig. 8 red is polish and blue is western, and in Fig. 9 its the reverse. This isn't essential, but would look better if they were consistent throughout both figs. \larry{fixed. Also, Fig. 8 was black for Polish, I changed it to blue so the two figures are in agreement.}
\label{pair_freq}
\end{figure}

It has previously been noted that the frequency of occurrence of individual branches in $N-2$ malignancies is heavy-tailed~\cite{eppstein2012random, yang2017small}. A similarly heavy-tailed distribution is apparent in the distribution of occurrences of specific branch pairs in $N-3$ malignancies, in both the Polish and Western US (Fig.~\ref{pair_freq}).

\begin{figure}[!b]
\centering
\includegraphics[width=3.1in]{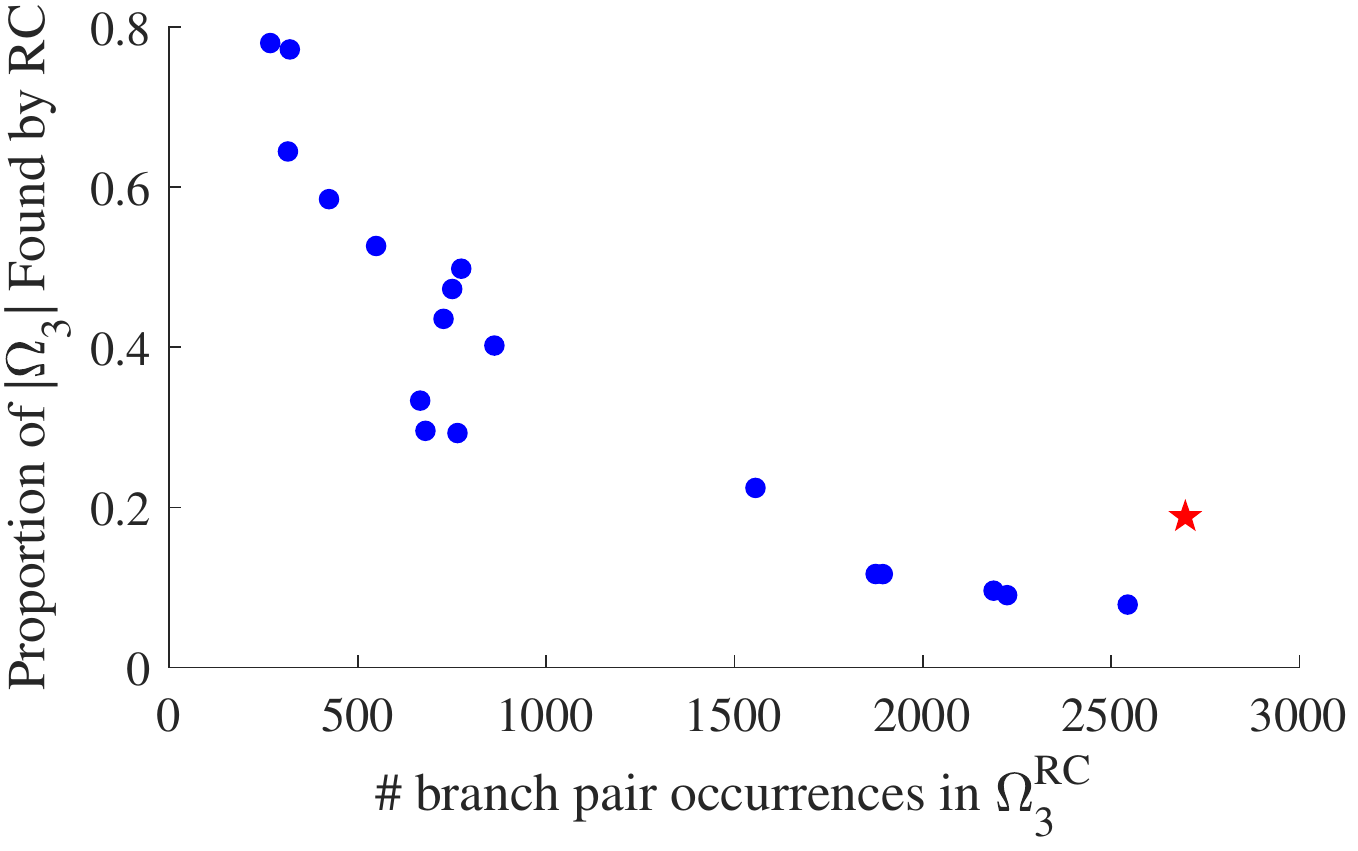}
\caption{The relation between number of occurrences of specific branch pairs in $N-3$ malignancies found by RC ($x$-axis) and the proportion RC has found of all $N-3$ malignancies that include those pairs ($y$-axis), for the Western US test case. Only the 20 most frequently occurring branch pairs are shown, with the star indicating the branch pair that occurred most frequently in $\Omega_3^{RC}$ ($Pair_{max})$.}
\label{systematic_bias}
\end{figure}

Further examination reveals that the set reduction scheme used in RC does, indeed, introduce a sampling bias when sampling from such heavy-tailed distributions. Specifically, the branch pairs that appear in disproportionately large numbers of $N-3$ malignancies are systematically under-sampled by the RC algorithm. To illustrate this, the 20 most frequently occurring branch pairs found in $|\Omega^{RC}_3|$ for the Western US test case were selected, and a brute force search of all possible $N-3$ contingencies that included each of these top 20 branch pairs (requiring $O(N)$ computation time for each branch pair) was performed. In Fig.~\ref{systematic_bias}, the proportion of $N-3$ malignancies found by RC that contain each of these branch pairs is plotted as a function of the observed number of occurrences of the branch pairs in $\Omega^{RC}_3$. A clear negative trend is present, with $N-3$ malignancies containing the most frequently occurring branch pairs severely under-sampled relative to $N-3$ malignancies containing less frequently occurring branch pairs. While a thorough explanation of the causes of this sampling bias are beyond the scope of this paper, here the bias is exploited to estimate both lower and upper bounds on $|\Omega_{3}|$.  (It is worth noting that, as $|\Omega^{RC}_3|$ approaches $|\Omega_3|$, the sampling bias of branch pairs found in $N-3$ malignancies decreases. However, for large networks such as the Western US test case, it is not computationally feasible to sample more than a small fraction of the $N-3$ malignancies, so the sampling bias remains high.)

Given that sampling probabilities are unequal, this problem is analogous to the common conservation biology task of estimating population sizes {\em via} mark-and-recapture surveys in closed populations with heterogeneous sampling probabilities. There are numerous techniques that have been developed for this kind of problem~\cite{amstrup2010handbook}. Here, Chao's method~\cite{chao1987estimating} is used, because it is known to be particularly robust to heterogeneous sampling probabilities. In the power system context, the ``population'' under consideration is $\Omega_k$, the set of all $N-k$ malignancies. Chao's estimate is calculated as $|\Omega_3|^{Chao}=|\Omega_3^{RC}| + n_1^2 / (2n_2)$, where $n_1$ is the number of $N-3$ malignancies found exactly once by RC sampling and  $n_2$ is the number of $N-3$ malignancies found exactly twice by RC sampling. Chao's method produces a lower-bound on the population size within a fixed confidence interval~\cite{chao1987estimating}, so it is assumed that $|\Omega_3|^{Chao} \leq |\Omega_3|$.\par

An upper-bound on $|\Omega_3|$ can be estimated by taking advantage of the two observations demonstrated above: (i) certain branch pairs appear disproportionately often in $N-3$ malignancies (Fig.~\ref{pair_freq}), and (ii) the most frequently occurring branch pairs are under-sampled, relative to less frequent branch pairs (Fig.~\ref{systematic_bias}). Based on these observations, the Random Chemistry Proportional (RCP) method is proposed as a way to estimate an upper bound on $|\Omega_3|$, as follows: (i) apply RC sampling for a sufficient number of trials such that the identity of the most frequently occurring branch-pair ($Pair_{max}$) in the $N-3$ malignancies of the growing set $\Omega^{RC}_3$ becomes stable (for the Western US test case, $Pair_{max}$, indicated by the star in Fig.~\ref{systematic_bias}, became stable after about 7000 RC trials), (ii) perform a brute force search of all possible $N-3$ contingencies that include $Pair_{max}$ (this requires only $O(N)$ simulations) to determine the true number of $N-3$ malignancies that include $Pair_{max}$, (iii) compute what proportion of all $N-3$ malignancies that include $Pair_{max}$ were found by RC sampling, and finally (iv) we estimate the total number of $N-3$ malignancies (referred to as $|\Omega_3|^{RCP}$) by assuming that all other less-frequently occurring branch-pairs have found this same proportion of the total number of $N-3$ malignancies in which they occur.  Assuming that $Pair_{max}$ is under-sampled, this method provides an overestimate, and hence an upper-bound, on $|\Omega_3|$; i.e., it is expected that $|\Omega_3|^{RCP} > |\Omega_3|$.

\begin{figure}[!ht]
\centering
\includegraphics[width=3.2 in]{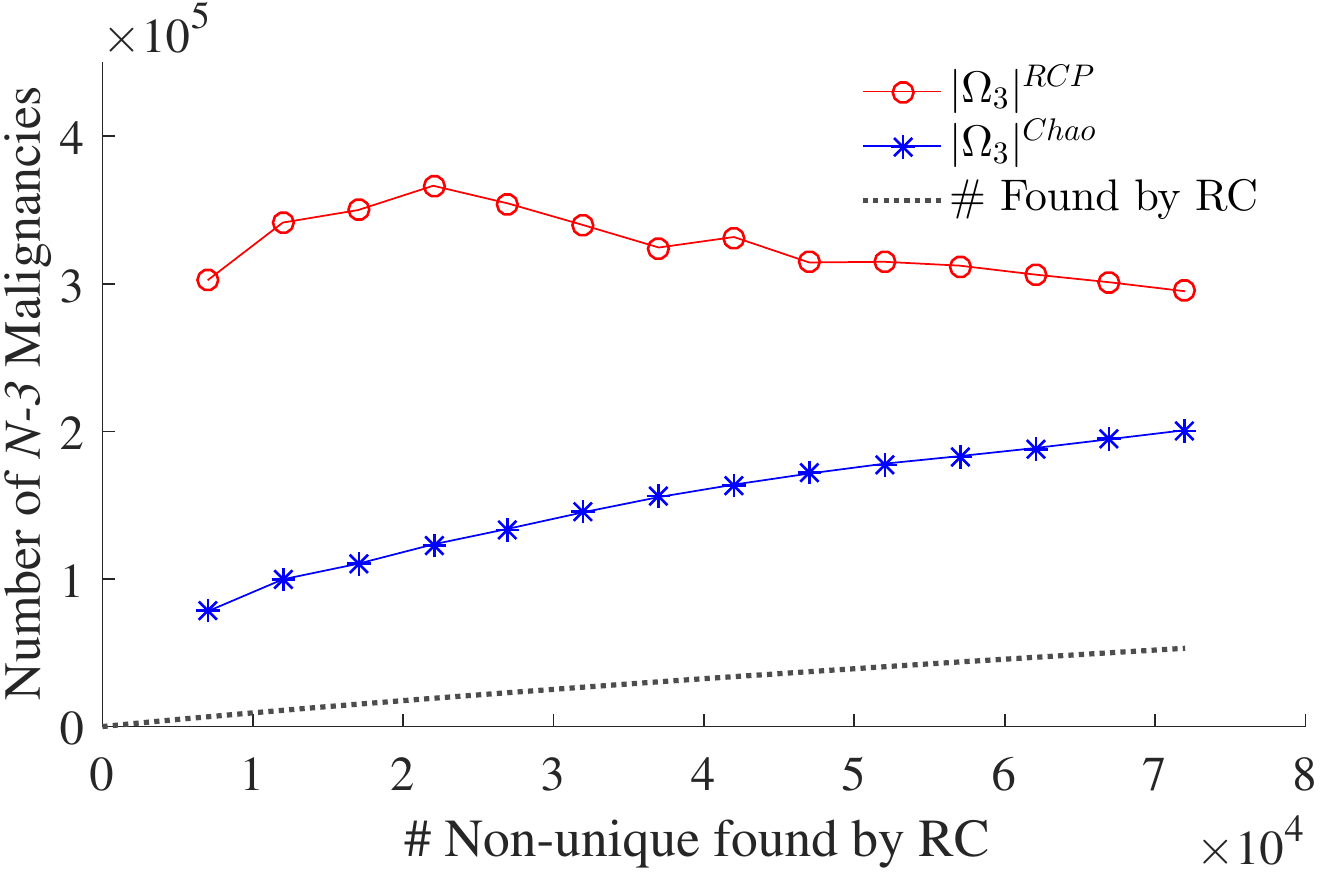}
\caption{The accumulation curve of $N-3$ malignancies found by RC is displayed below the lower-bound (Chao's method) and upper-bound (RCP method) estimates of $|\Omega_3|$ for the Western US test case.}
\label{BoundsFig}
\end{figure}

As the number of $N-3$ malignancies found by RC sampling increases, $|\Omega_3|^{Chao}$ and $|\Omega_3|^{RCP}$ can be seen to be converging (Fig.~\ref{BoundsFig}), thus increasing the confidence in these as lower and upper bounds on the true value of $|\Omega_3|$. Risk estimates are calculated for the rightmost values of $|\Omega_3|^{Chao}$ and $|\Omega_3|^{RCP}$ shown in Fig.~\ref{BoundsFig}, to obtain approximate bounds on risk due to $N-3$ malignancies for the Western US test case.

% In principle, these approaches to estimating lower and upper bounds can also be applied for estimating $|\Omega_k|$ for $k>3$, as long as $\Omega^{RC}_k$ is sufficiently large such that the $(k-1)$-tuple that is most frequent in $N-k$ malignancies found by RC sampling has stabilized.

Similar approaches could conceivably be applied for estimating $|\Omega_k|$ for $k>3$, however in this study $|\Omega^{RC}_4|$ and $|\Omega^{RC}_5|$ were insufficient to support this.

\section{Results} \label{sec:results}

\subsection{Set sizes of $\Omega_2$ and $\Omega_3$}\label{setsizes}
Brute force search was used to verify that RC sampling found all $N-2$ malignancies in the Polish base test case, with $|\Omega_2| = 540$. It is assumed that RC sampling also found all $N-2$ malignancies in the Western US test case with $|\Omega_2|=564$, since the accumulation curve became flat (Fig.~\ref{accum_curves}(top)). Using the non-linear curve-fitting method of~\cite{rezaei2015estimating},
$|\Omega_3|$ is estimated to be $\approx 6.4 \times 10^4$ in the Polish test case at the base load.
Using the Chao lower-bounding method~\cite{chao1987estimating} and the RCP upper-bounding method (described in Section~\ref{sec:num_of_maligs}), it is estimated that $2.0\times 10^5 \le |\Omega_3| < 2.9\times 10^5$ in the Western US test case.

\subsection{Impact of $N-2$ Correlation and Load Level on Risk} \label{load_level_results}

As shown in prior work~\cite{rezaei2015rapid, rezaei2015estimating}, the load levels on the Polish grid can greatly affect the vulnerability of the network to cascading power failure due to $N-2$ malignancies. As noted in~\cite{rezaei2015estimating}, risk varies non-linearly and non-monotonically with load, in part due to variations in the proximity of generation to demand that result from optimal power flow dispatch at different load levels. Risk actually tends to drop at very high load levels because the security constrained optimal power flow results in more local generation,  thus reducing the flow on critical long-distance transmission lines that can participate in many $N-k$ malignancies when overloaded. For a direct comparison to the results presented in~\cite{rezaei2015estimating}, the impact of spatial correlation in $N-2$ malignancies on risk was assessed as a function of load in the Polish test case.

\begin{figure}[!t]
\centering
\includegraphics[width=3.3in]{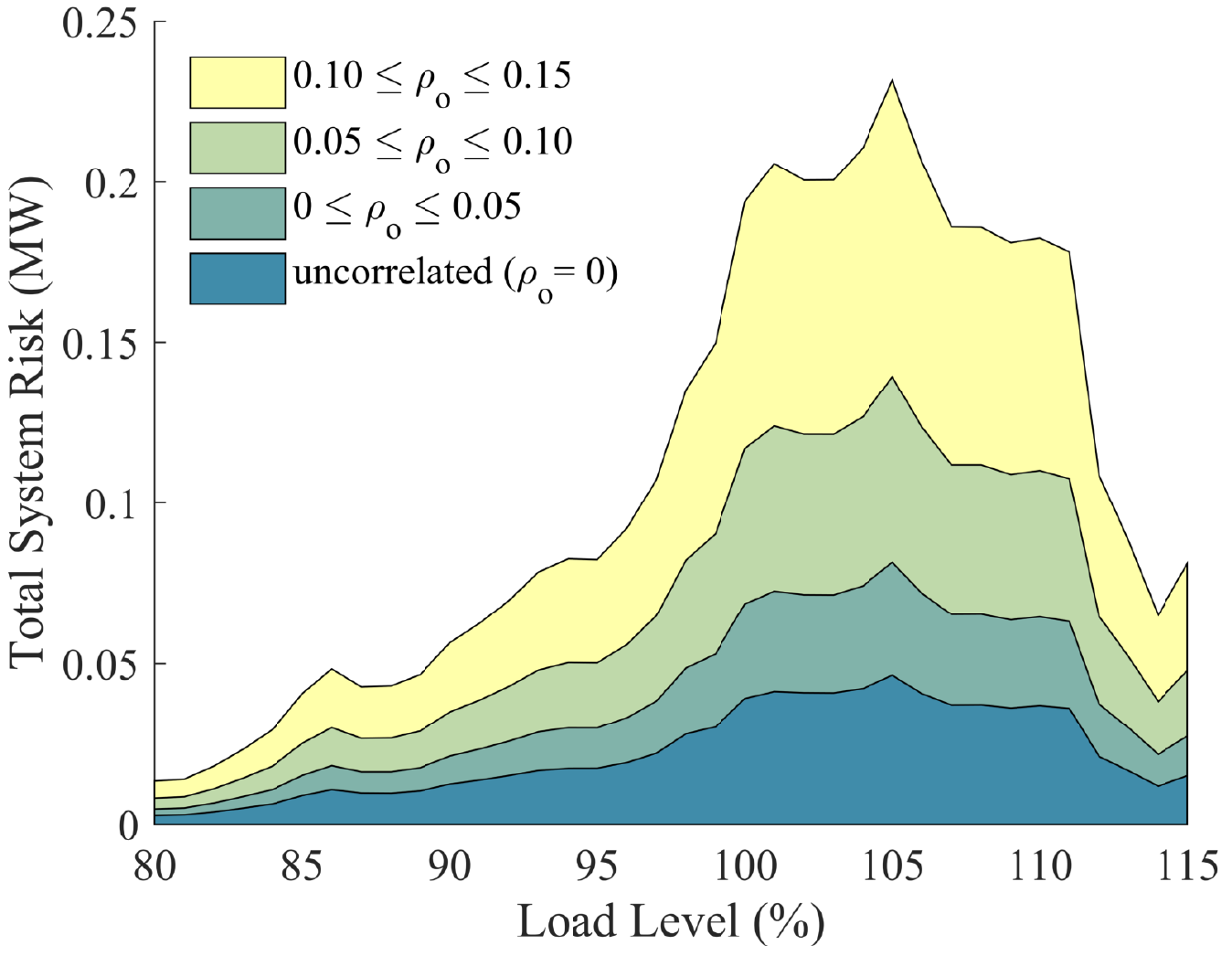}
\caption{Risk of cascading blackouts posed by spatially-correlated $N-2$ malignancies with a fixed characteristic length ($L=300$ km) and varying values of maximum correlation $\rho_o$ for load levels from 80\%-115\% of the Polish base test case.}
\label{results_rho_o}
\end{figure}

Changes in the system risk as a function of load at $L=300$ km (the longest characteristic correlation length tested) for 3 values of $\rho_o$  are illustrated in Fig.~\ref{results_rho_o}. Risk increased faster than linearly as a function of linearly increasing $\rho_o$, at each of the given load levels. The largest percentage increase in system risk occurred at load level 114\%, while the greatest absolute increases in risk occur between load levels of 97\%-112\%, where there are the most $N-2$ malignancies. In general, while introducing correlation in initiating outages magnifies the risk of cascading blackouts, it does not fundamentally alter the overall shape of the risk curve as a function of load at $L=300$ km. \par

\begin{figure}[!t]
\centering
\includegraphics[width=3.3in]{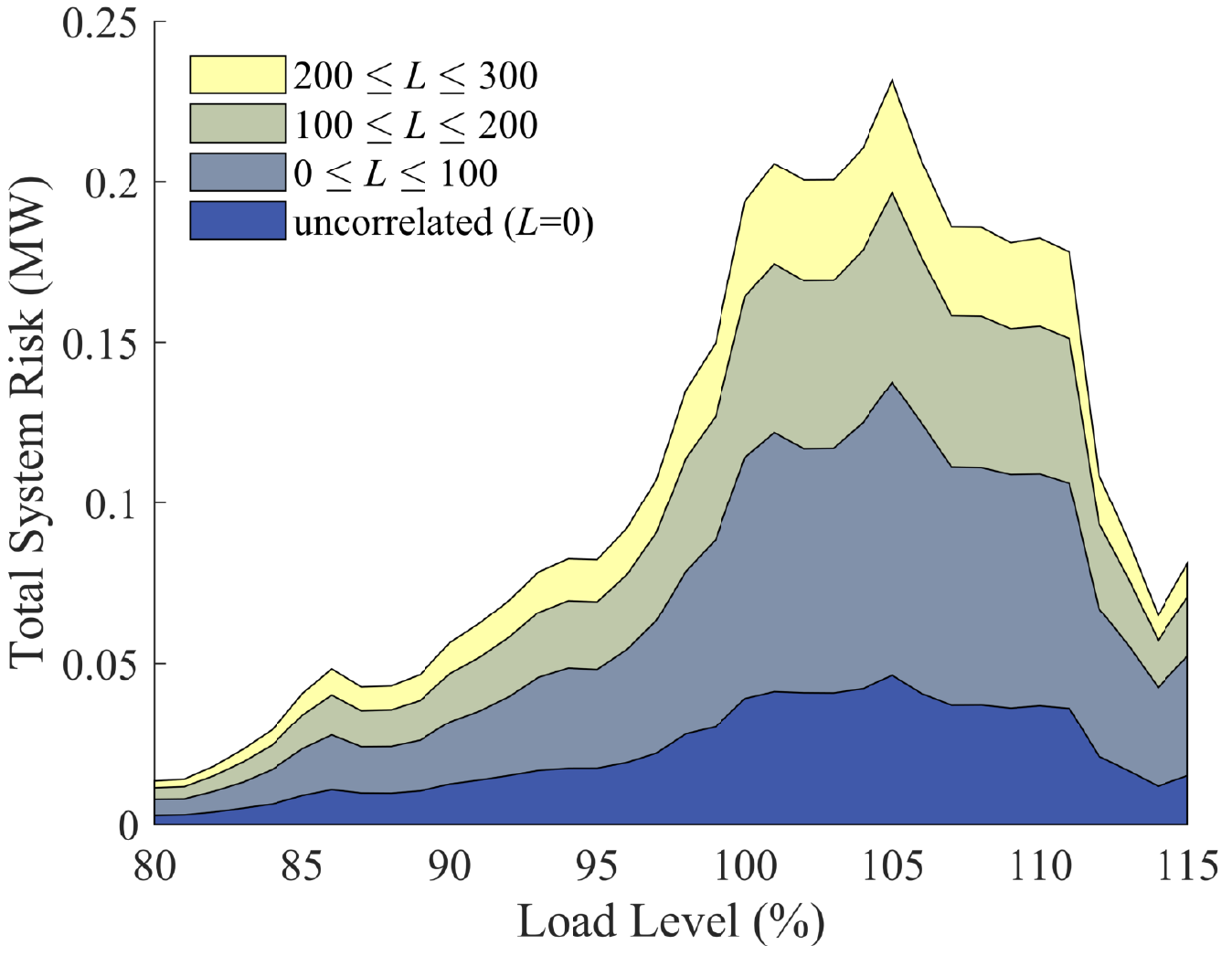}
\caption{Risk of cascading blackouts posed by spatially-correlated $N-2$ malignancies with a fixed maximum correlation ($\rho_o=0.15$) and varying values of characteristic length $L$ (in km) for load levels from 80\%-115\% of the Polish base test case.}
\label{results_L}
\vspace{-12pt}
\end{figure}

When $\rho_o=0.15$ (the largest $\rho_o$ tested) and $L$ was varied from 0 to 300 km, results were superficially similar to those in Fig.~\ref{results_rho_o}, in that higher correlation increases risk without changing the overall shape of the risk curve as a function of load. As was the case when $L$ was fixed, the largest percentage increase in system risk was found to occur at load level 114\%, and greatest absolute increases in risk occured between load levels of 97\%-112\%.  However, in this case risk increases slower than linearly with linear increases in $L$, with the largest increases occurring for intermediate values of $L$ (Fig.~\ref{results_L}). This occurs because increasing $L$ beyond a certain point has diminishing impact on correlation, as $L$ approaches the radius of the network.\par

\begin{figure}[!t]
\centering
\vspace{.1in}
\includegraphics[width=3.25in]{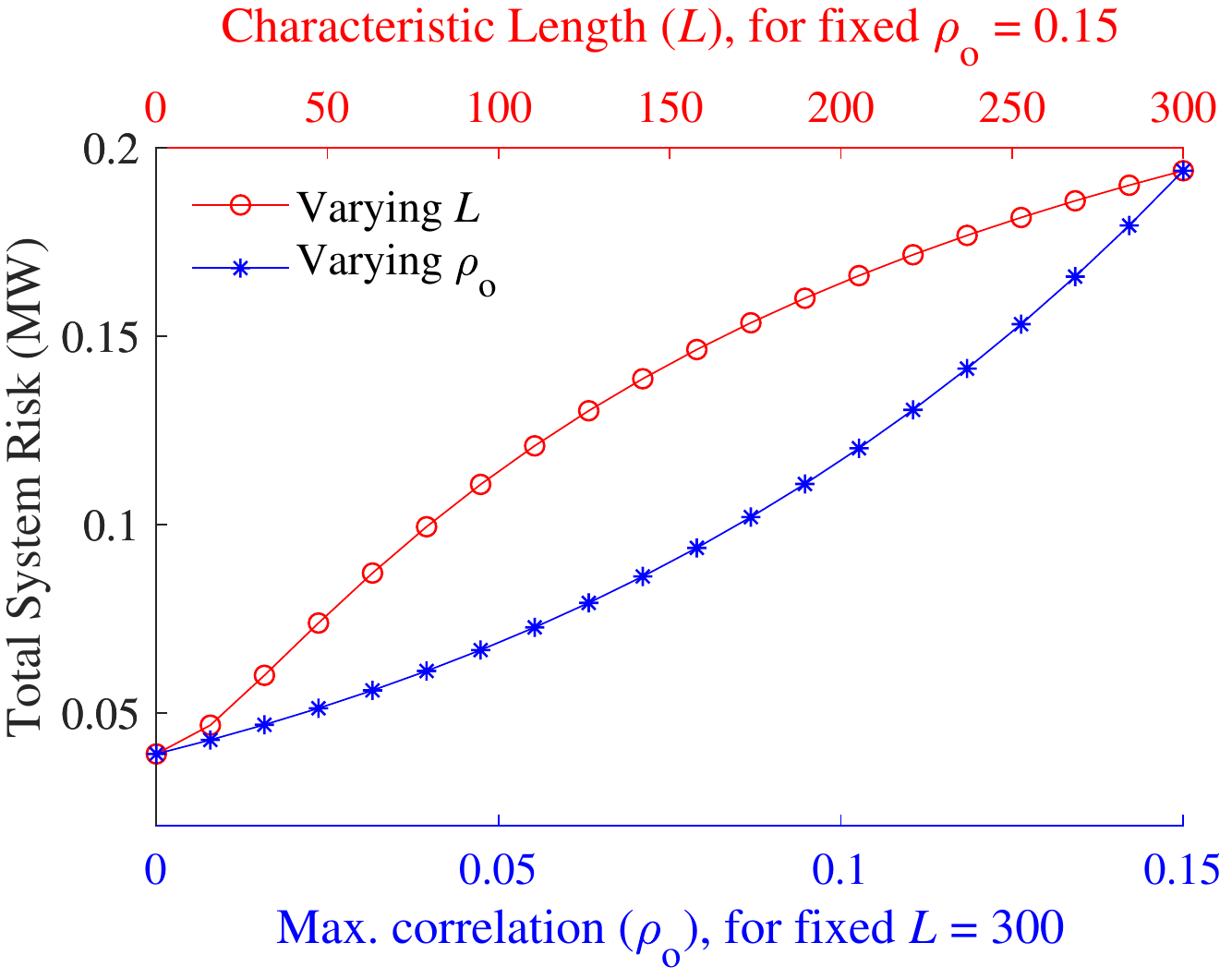}
\vspace{-.2in}
\caption{Comparing change in risk of cascading blackouts for varying $L$ (in km) with $\rho_o$ fixed at 0.15 (top $x$-axis) {\em vs.} varying $\rho_o$ with $L$ fixed at 300 km (bottom $x$-axis) for the Polish base test case.}
\label{L_vs_rho_o}
\end{figure}

 The super-linear increases in risk as a function of $\rho_o$ and sub-linear increases in risk as a function of $L$ at the base load are clearly illustrated in Fig.~\ref{L_vs_rho_o}.

\subsection{Risk from $N-2$ and $N-3$ Malignancies}

Risk of cascading blackouts posed by $N-2$ and $N-3$ malignancies was computed for both the Polish base test case and the Western US test case, over all values of $L$ and $\rho_o$ tested, using the set size estimates given in Sec.~\ref{setsizes}.   \par

For the Polish test case (Table~\ref{polish_results_2s_and_3s}), the increase in estimated risk due to spatial correlation ranges from 149\% for the most modest level of correlation tested ($L=100$ km, $\rho_o=0.05$) to 582\% in the most extreme case tested ($L=300$ km, $\rho_o=0.15$), relative to the uncorrelated case.

\begin{table}[!hb]
%% increase table row spacing, adjust to taste
\renewcommand{\arraystretch}{1.3} % Heightens rows
\setlength{\tabcolsep}{10pt} % Widens columns
\caption{Risk Attributable to $N-2$ and $N-3$ Malignancies in the Polish Test Case for Varying Levels of Spatial Correlation. }
\label{polish_results_2s_and_3s}
\begin{center}
\rowcolors{1}{white}{light-gray}
\begin{tabular}{ ccccc }
\hline
  & \multicolumn{4}{c}{$L$ (km)} \\ \cline{2-5}
  $\rho_o$ & 0 & 100 & 200 & 300 \\ \hline
  0.00 & 0.0394 & - & - & - \\
  0.05 & - & 0.0586 & 0.0675 & 0.0720 \\
  0.10 & - & 0.0876 & 0.1142 & 0.1290 \\
  0.15 & - & 0.1314 & 0.1918 & 0.2293 \\
  \hline
\end{tabular}
\end{center}
\end{table}

\begin{table}[!t]
%% increase table row spacing, adjust to taste
\renewcommand{\arraystretch}{1.3} % Heightens rows
\setlength{\tabcolsep}{10pt} % Widens columns
\caption{Estimated Lower Bounds ($LB$) and Upper Bounds ($UB$) on Risk Attributable to $N-2$ and $N-3$ Malignancies in the Western US Test Case for Varying Levels of Correlation.}
\label{westernUS_results_2s_and_3s}
\begin{center}
\rowcolors{1}{white}{light-gray}
\begin{tabular}{ cccccc }
\hline
  & & \multicolumn{4}{c}{$L$ (km)} \\ \cline{3-6}
  $\rho_o$ & & 0 & 100 & 200 & 300 \\ \hline
  0.00 &
    \begin{tabular}{@{}c@{}}$LB$ \\ $UB$\end{tabular} &
    \begin{tabular}{@{}c@{}}0.0654 \\ 0.0665\end{tabular}
    & - & - & - \\
  0.05 &
    \begin{tabular}{@{}c@{}}$LB$ \\ $UB$\end{tabular} & - &
    \begin{tabular}{@{}c@{}}0.0846 \\ 0.0864\end{tabular} &
    \begin{tabular}{@{}c@{}}0.0950 \\ 0.0974\end{tabular} &
    \begin{tabular}{@{}c@{}}0.1019 \\ 0.1048\end{tabular} \\
  0.10 &
    \begin{tabular}{@{}c@{}}$LB$ \\ $UB$\end{tabular} & - &
    \begin{tabular}{@{}c@{}}0.1148 \\ 0.1181\end{tabular} &
    \begin{tabular}{@{}c@{}}0.1444 \\ 0.1502\end{tabular} &
    \begin{tabular}{@{}c@{}}0.1654 \\ 0.1735\end{tabular} \\
  0.15 &
    \begin{tabular}{@{}c@{}}$LB$ \\ $UB$\end{tabular} & - &
    \begin{tabular}{@{}c@{}}0.1631 \\ 0.1701\end{tabular} &
    \begin{tabular}{@{}c@{}}0.2293 \\ 0.2445\end{tabular} &
    \begin{tabular}{@{}c@{}}0.2801 \\ 0.3036\end{tabular} \\
  \hline
\end{tabular}
\end{center}
\end{table}

For the Western US test case, (Table~\ref{westernUS_results_2s_and_3s}), the increase in lower (upper) bounds on risk estimates varied from 129\% (130\%) for the most modest level of correlation tested ($L=100$ km, $\rho_o=0.05$) to 428\% (456)\% in the most extreme case tested ($L=300$ km, $\rho_o=0.15$), relative to the uncorrelated case.

For both test cases, the general effect of $L$ and $\rho_o$ on risk that is described in Section~\ref{load_level_results} also holds in these results. That is, risk tends to grow faster than linearly with respect to $\rho_o$ and slower than linearly with respect to $L$.  The larger proportionate increases in the Polish test case, relative to the Western US test case, occur because the average distance between branches in malignancies are shorter than in the Western US test case (Fig.~\ref{malig_dist}), thus magnifying the impacts of spatial correlation.

\subsection{Relative Risk of $\boldsymbol{N-2}$ {\em vs.} $\boldsymbol{N-3}$ Malignancies}

%Specifically, for a given pair and triplet of branches, referred to here as $\sigma_2$ and $\sigma_3$, respectively, assume each branch has independent failure probability $p$. When $\rho = 0$ (independent outages), we find the joint probability of failures to be $P(\sigma_2) = p^2$ and $P(\sigma_3) = p^3$. When correlation is at a maximum $\rho = 1$, if any one branch fails, all will fail. In this scenario, $P(\sigma_2) = P(\sigma_3) = p$. Hence, it is necessary for the joint probability of the branches in $\sigma_3$ failing to grow more rapidly with correlation than the probability of the branches in $\sigma_2$ failing. This will cause the relative risk of $N-3$ malignancies to grow compared to $N-2$ malignancies, making consideration of these higher-order malignancies that much more important. \maggie{I'm finding this paragraph rather convoluted, and perhaps not necessary?} \par

It is expected that $N-3$ malignancies will contribute more to risk when there is spatial correlation in initiating outages, but it is not clear to what degree. There are several factors that could potentially disproportionately affect the impact of $N-3$ malignancies on risk when there is spatial correlation, relative to that of $N-2$ malignancies, including: (i) size of blackouts caused by $N-3$ {\em vs.} $N-2$ malignancies; (ii) the independent probability of branch outages in $N-3$ {\em vs.} $N-2$ malignancies; (iii) the distance between branches in $N-3$ {\em vs.} $N-2$ malignancies. These factors are each discussed in more detail below.

\subsubsection{Blackout Sizes}
\begin{figure}[!ht]
\centering
\includegraphics[width=3.6in]{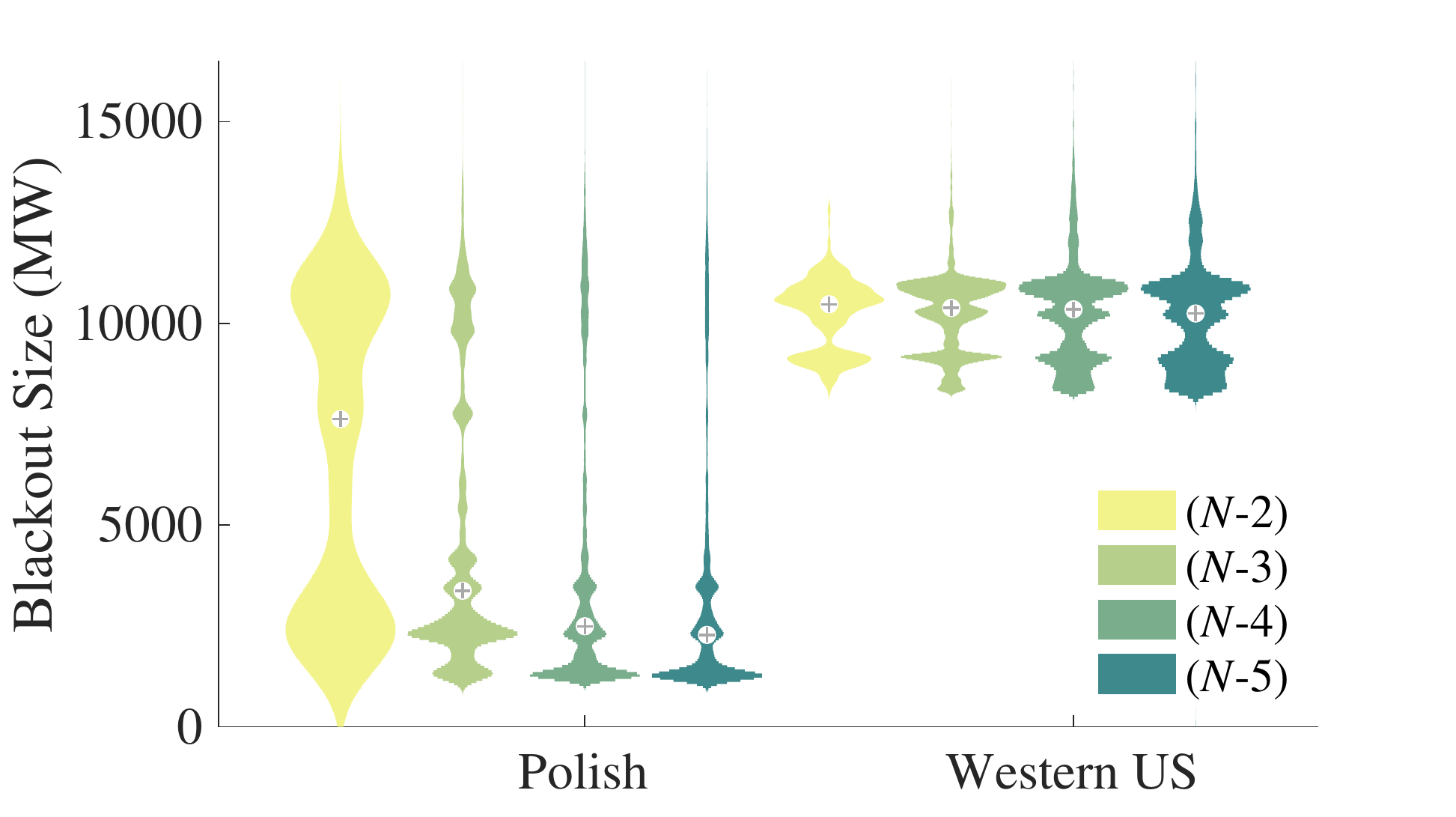}
\caption{Distributions of blackout sizes (in total MW load shed) caused by all $N-k$ malignancies ($2\leq k \leq 5$) found by RC sampling, for the Polish and Western US test cases. For clarity, medians are marked with crosshairs and each distribution has been independently normalized to the same maximum width.}
\label{bo_sizes_2_vs_3}
\end{figure}

If the sizes of blackouts resulting from $N-3$ malignancies were larger than $N-2$ malignancies, this could disproportionately increase the relative contribution of $N-3$ malignancies to risk when spatial correlation is present. However, we have observed that the sizes of cascading blackouts tend to follow similarly shaped distributions, independent of the number of component outages in the triggering event, due to similar patterns of network separation. This is illustrated by the distributions of blackout sizes (as estimated by DCSIMSEP) caused by all $N-2$ malignancies, and the subsets of identified $N-3$ malignancies found by RC sampling, for both the Polish and Western US test cases (Fig.~\ref{bo_sizes_2_vs_3}). In both test cases, the median blackout size for the identified $N-3$ malignancies was actually lower than those caused by the $N-2$ malignancies. Specifically, for the Polish test case, the median blackout size caused by $N-2$ malignancies was 7,624 MW {\em vs.} 3,372 MW for those caused by identified $N-3$ malignancies. In the Western US test case, the median blackout size from $N-2$ malignancies was 10,473 MW whereas from the $N-3$ malignancies it was 10,382 MW.  These patterns and trends continue in the identified sets of $N-4$ and $N-5$ malignancies (Fig.~\ref{bo_sizes_2_vs_3}).

\subsubsection{Independent Branch Outage Rates}
In this study, independent outage rates were assumed to be homogeneous for all branches. However, in a real system the distribution of independent outage rates will be heterogeneous. If branches that are typically involved in $N-3$ malignancies are independently more likely to fail than those involved in $N-2$ malignancies, this could inflate the relative risk of $N-3$ malignancies when spatial correlation is present. While there is no obvious rationale for why this might be true, the observation that branches that occur frequently in $N-2$ malignancies also appear frequently in $N-3$ malignancies is indirect evidence against this. For example, in the Polish network, 8 of the 10 most frequently occurring branches in $N-2$ and $N-3$ malignancies are shared, accounting for 44\% and 24\% of all $N-2$ and $N-3$ malignancies found, respectively. Likewise, for the Western US test case, 9 of the 10 most frequently occurring branches in $N-2$ malignancies are also in the top 10 most frequently occurring $N-3$ malignancies, accounting for 49\% and 29\% of all $N-2$ and $N-3$ malignancies, respectively.

\subsubsection{Distance Between Branches}
\begin{figure}[!ht]
\centering
\includegraphics[width=3.6in]{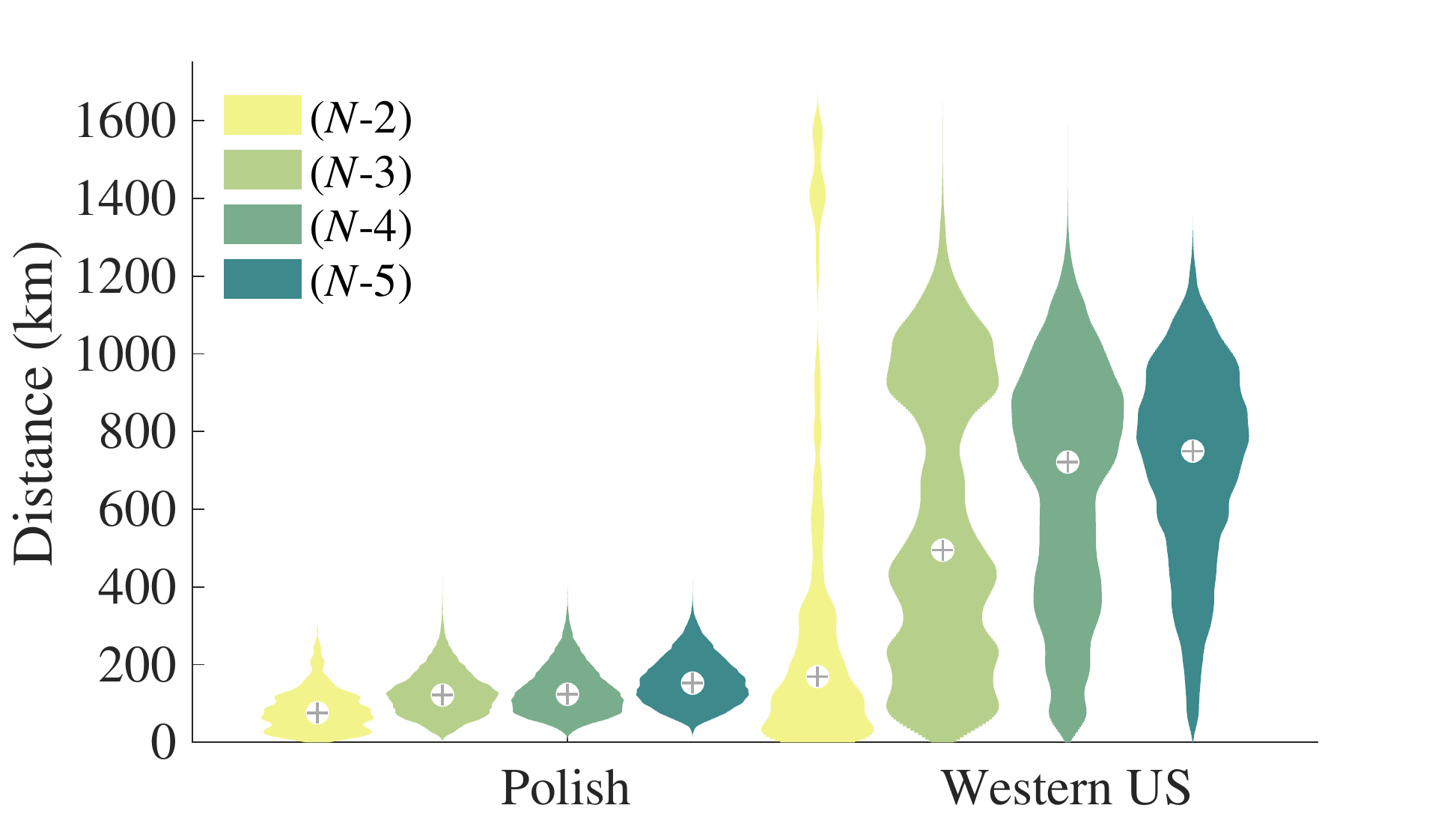}
\caption{Distributions of pairwise distances among branches in all $N-k$ malignancies ($2 \leq k \leq 5$) identified by RC sampling, for the Polish and Western US test cases. For clarity, medians are marked with crosshairs and each distribution has been independently normalized to the same maximum width.}
\label{dist_2_vs_3}
\end{figure}

The distance between branches in $N-3$ {\em vs.} $N-2$ malignancies will obviously impact the degree to which spatial correlation will increase their relative contributions to risk. In both the Polish and Western US test cases,  median distances between all pairs of branches occurring in identified $N-k$ malignancies increases with $k \in \{2,3,4,5\}$ (Fig.~\ref{dist_2_vs_3}). Specifically, in the Polish test case the median distances between pairs of branches were 76.1 km and 121.8 km, in $N-2$ and $N-3$ malignancies, respectively; in the Western US test case the medians were 169.4 km and 494.6 km in the $N-2$ and $N-3$ malignancies, respectively. This helps to mitigate the increase in relative risk from $N-3$ {\em vs.} $N-2$ malignancies that occurs as a result of spatial correlation.

\subsubsection{Comparing Relative Risk with Correlation}
\begin{figure}[!ht]
\centering
\includegraphics[width=3in]{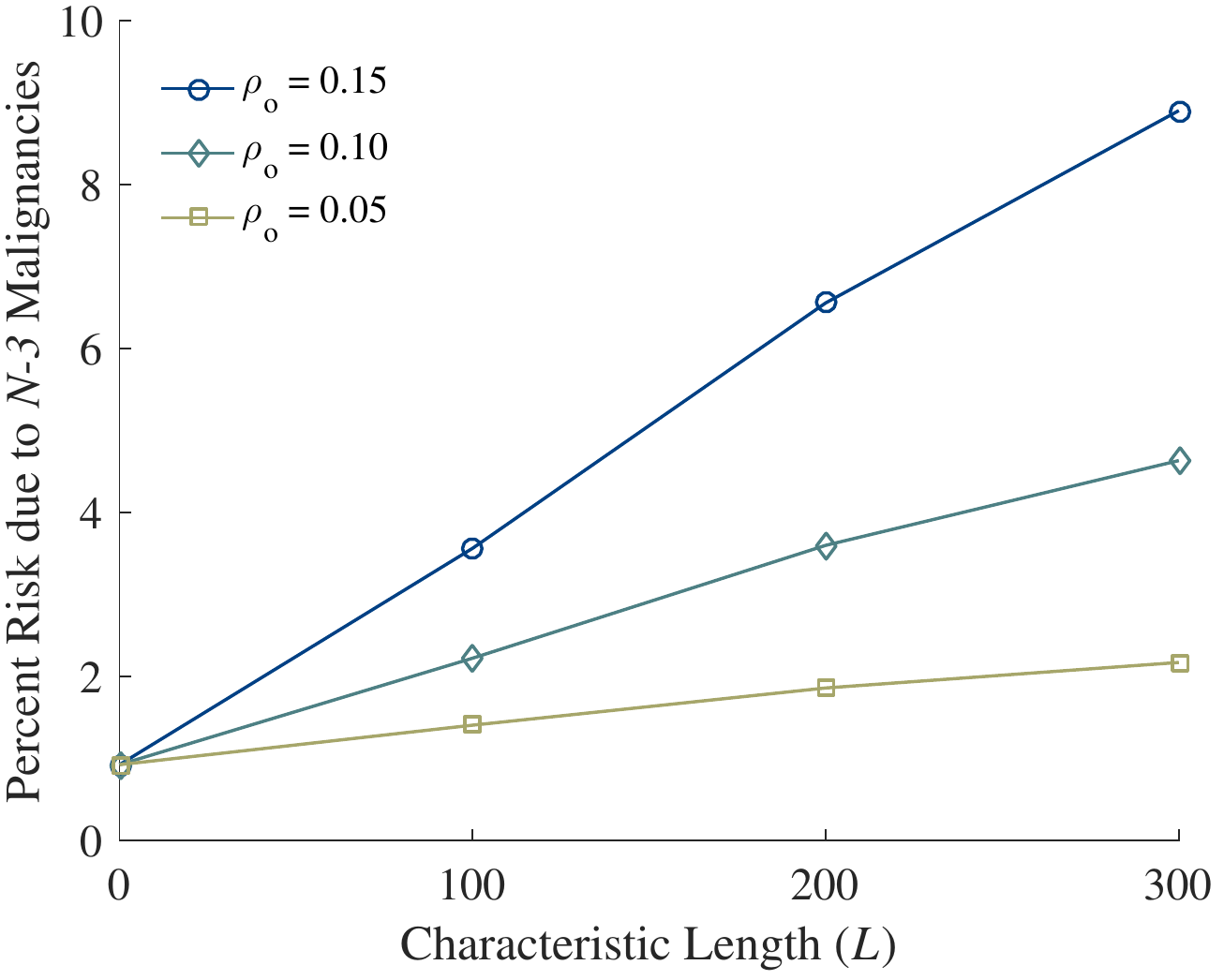}
\caption{Estimated percentage of risk attributable to $N-3$ malignancies {\em vs.} $N-2$ malignancies for the Polish Test Case under varying levels of correlation, including all combinations of $L \in \{0,100,200,300\}$ km and $\rho_o \in \{0,0.05,0.10,0.15\}$.}
\label{polish_2_vs_3_results}
\vspace{-.2in}
\end{figure}

\begin{figure}[!ht]
\centering
\includegraphics[width=3.3in]{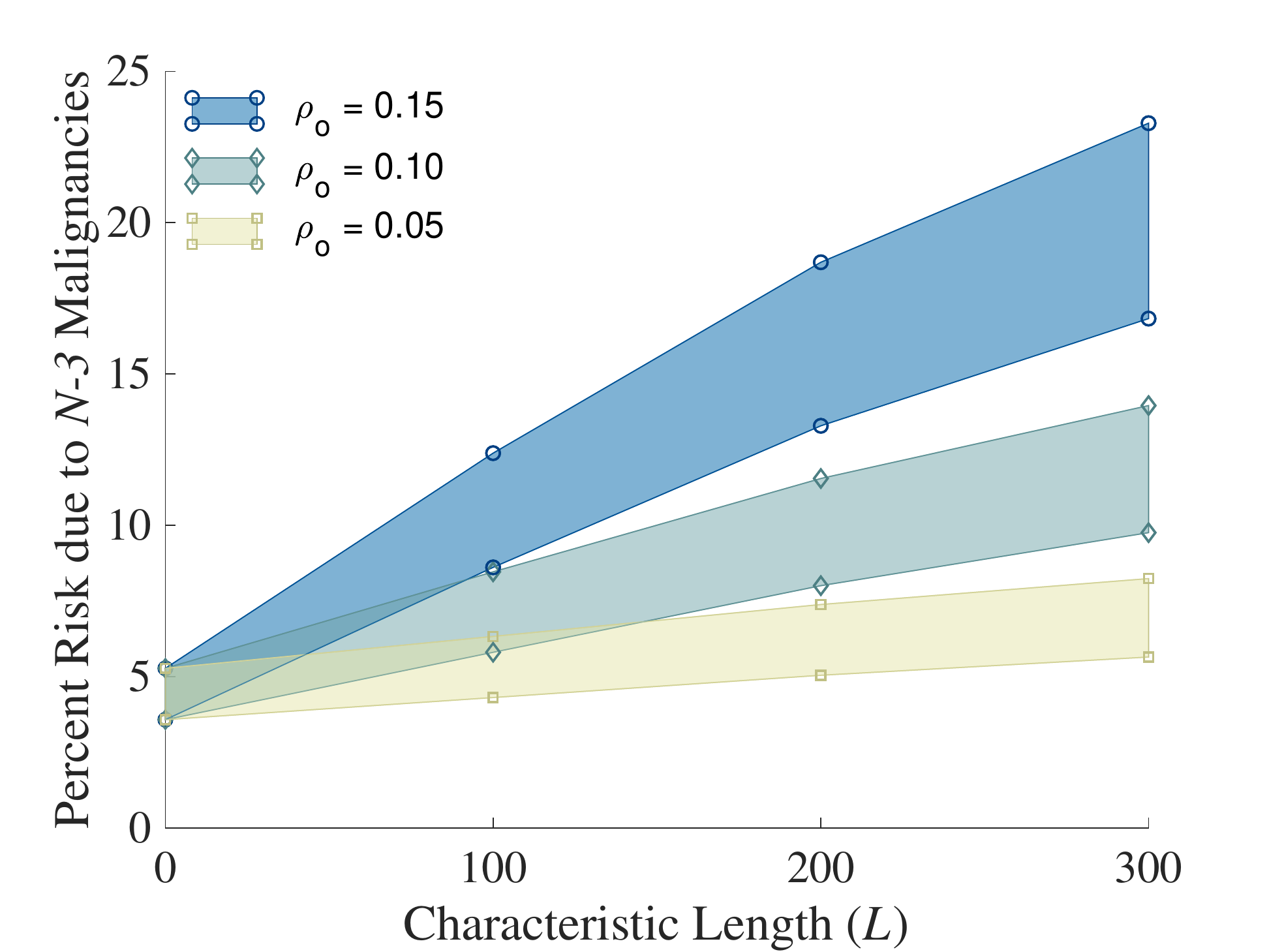}
\caption{Shaded regions represent bounded estimates on the percentage of risk attributable to $N-3$ malignancies {\em vs.} $N-2$ malignancies for the Western US Test Case under varying levels of correlation, including all combinations of $L \in \{0, 100,200,300\}$ km and $\rho_o \in \{0, 0.05,0.10,0.15\}$.}
\label{westernUS_2_vs_3_results}
\end{figure}

For the Polish test case, $< 1\%$ of risk can be attributed to $N-3$ malignancies when there is no correlation whereas under the highest level of correlation considered ($L=300$  km, $\rho_o=0.15$), the share of risk associated with $N-3$ malignancies rises to around 9\% (Fig.~\ref{polish_2_vs_3_results}). Similarly, for the Western US test case, $N-3$ malignancies account for 3\%-5\% of risk when there is no correlation, but between 16\%-24\% under the maximal correlation  ($L=300$ km, $\rho_o=0.15$) considered (Fig.~\ref{westernUS_2_vs_3_results}). \par

\section{Discussion} \label{sec:discussion}

% [IN DISCUSSION, MENTION THE LACK OF NECCESARY DATA FOR UNDERSTANDING THE UNDERLYING DISTRIBUTIONS WE’RE TRYING TO MODEL, AND HOW THERE IS INCREASING EFFORT AND PROGRESS IN THIS REALM, SEE pacmeieee, papic2017research, REF2, REF3]. \par

%Much of the past work in quantifying risk to power systems from cascading blackouts assumed independence of triggering component outages (i.e., no correlation).

%Due to the challenges associated with predicting the number of higher-order malignancies, it is important to consider the relative risk associated with such rare events.
Previous research into cascading failure risk %< minor edit for those who jump straight to the conclusions
demonstrated that $N-3$ malignancies constitute a relatively low proportion of risk compared to $N-2$ malignancies, assuming initiating branch outage independence~\cite{rezaei2015estimating}. This suggests that, if initiating outages are generally caused by independent events, limiting risk analysis to the more computationally tractable $N-2$ malignancies may be sufficient to capture the majority of risk. However, in reality, common causes such as relay failures, weather disturbances, earthquakes, fire, or spatially localized terrorist attacks may trigger multiple near-simultaneous outages in geographic proximity that could potentially result in cascading blackouts. For these cases an assumption of independence will under-estimate cascading blackout risk.

This paper presents a method that uses %< minor edit
copula analysis as a flexible, customizable approach for incorporating correlation into risk calculations, building on preliminary work presented in~\cite{clarfeld2018assessing}. The impact of spatial correlation in $N-2$ and $N-3$ initiating outages on risk of cascading blackouts is assessed in the Polish test case as well as the much larger, and geographically more realistic, Western US test case. \par

The Western US test case has over four times as many branches as the Polish test case, and the increased computational cost of performing cascade simulations combined with the substantially larger search space of $N-3$ contingencies rendered previously developed methods~\cite{eppstein2012random, rezaei2015estimating} for estimating the set size of $N-3$ malignancies ineffective. Thus, extending the approach to include risk from $N-3$ malignancies in the Western US test case required new methods %< new methods, not new studies of methods
to estimate lower- and upper-bounds on the total number of $N-k$ malignancies, for $k=3$. \par

The results indicate that when spatial correlation is present in initiating outages, the relative contribution of $N-3$ malignancies to risk of cascading blackouts increases, although the increase is partially mitigated by the fact that pairwise distances between branches in $N-3$ malignancies are greater than in $N-2$ malignancies.

It is expected that the impact of even higher-order malignancies will similarly increase with increasing spatial correlation, even though median pairwise distances between branches in malignancies continue to increase with $k$. In principle, the approaches to estimating lower and upper bounds on $|\Omega_3|$ presented here for the Western US test case could also be applied for estimating $|\Omega_k|$ for $k>3$, given a sufficient number of RC trials. We are currently exploring these and other methods on large synthetic networks to establish their limitations as a function of network size and functional heterogeneity. While ignoring higher-order $N-k$ malignancies when component outages are correlated will likely underestimate the magnitude of the risk of cascading failures, estimating $|\Omega_k|$ for $k$ higher than 3 (or possibly 4) on large networks may not be computationally tractable, due to the sheer number of these high-order malignancies. Preliminary work indicates that, for the Western US test case, $|\Omega_4|$ may be at least two orders of magnitude higher than $|\Omega_3|$. \par
%However, we have also shown that incorporation of $N-3$ malignancies does not qualitatively alter the shape of the risk curve as a function of load level. This is encouraging, as it implies that an $N-2$ risk analysis may still be useful for planning purposes, to analyze how the shape of the risk curve changes in response to modifications to the grid.

The lack of accurate data regarding independent transmission line outage rates and the impacts of common cause events on those rates are an important limitation to applying these methods in practice. Some such data is available to industry through systems such as the NERC TADS database, but these data are typically unavailable for research purposes. Increasingly, efforts are being made to better predict how common-cause events such as weather-related events will impact the grid~\cite{liu2017risk}. Such knowledge will inform specific applications of the generalized framework introduced herein.  \par

The methods presented in this paper should work with any simulator (AC, DC, or even something more sophisticated like a full dynamics cascading failure simulator). However, more complicated simulation models require much larger input datasets and tuning these models to get accurate results is a longer process (see~\cite{song2016dynamic, kosterev1999model} for illustrations of the challenges associated with dynamic modeling of cascading failure). For example, in order to get an accurate model from an AC or dynamic power system simulator one would need to accurately model all of the dynamic reactive power elements, such as synchronous condensers and switched capacitor banks, in order to get accurate results. The impact of these controls is that a system with large amounts of reactive support will act more like a DC model with uniform voltage, than an AC model without reactive support. Since the focus of this paper is primarily on the computational method for risk analysis, rather than precise power systems details, we have used a simulator based on the DC power flow. Based on our experience with more complicated simulation models, we do not expect that a more complicated simulator would produce qualitative differences in the results, %< minor edit here
although quantitative differences would result since there are more mechanisms of cascading in an AC power flow model.

Future work will study the impact of parametric choices such as distance metrics, correlation functions, and the underlying branch outage probability distributions. The method will also be applied to study the risk of cascading failure in interdependent networks, which are ubiquitous in human-engineered infrastructures~\cite{rinaldi2001identifying}. For example, coupling between communication and power networks can substantially impact their robustness to cascading failures~\cite{korkali2017reducing, chen2018robustness}; the method presented herein could help to quantify the impact of this coupling on risk in the presence of correlated component outages. As new methods for measuring the risk of cascading failure in systems with correlated initiating event probabilities emerge, there will be a need for comparisons to better understand the relative computational efficiency and accuracy of these approaches.\par

\ifCLASSOPTIONcompsoc
  \section*{Acknowledgments}
\else
  \section*{Acknowledgment}
\fi

This work was supported in part by the National Science Foundation, award numbers ECCS-1254549, CNS-1735513 and DGE-1144388. The computational resources provided by the University of Vermont Advanced Computing Core (VACC) are gratefully acknowledged.

\ifCLASSOPTIONcaptionsoff
  \newpage
\fi

\vskip -2\baselineskip plus -1fil

\begin{IEEEbiography}[{\includegraphics[width=1in,height=1.25in,clip,keepaspectratio]{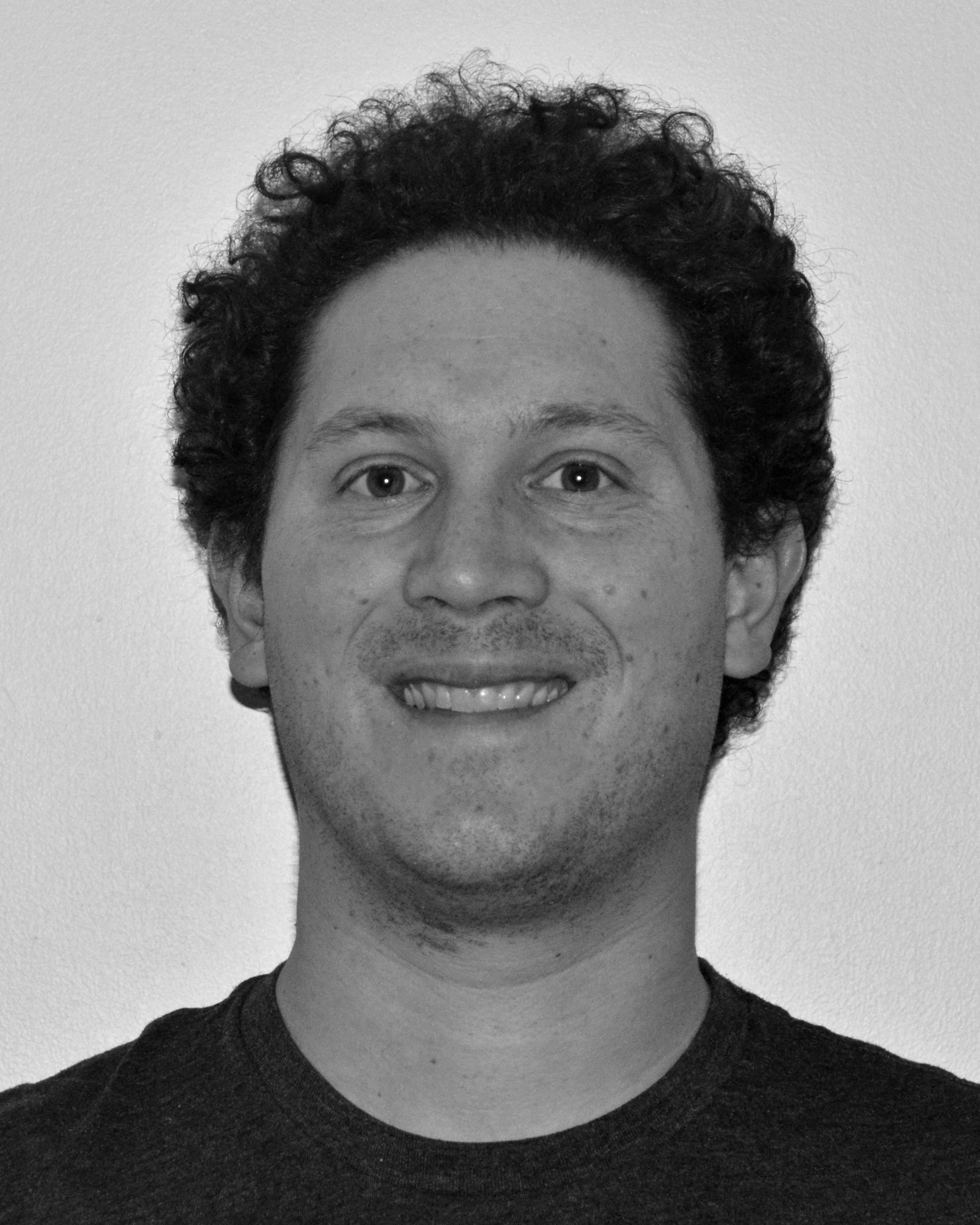}}]{Laurence A. Clarfeld}
 received his B.S.~in Mathematics from the University of Vermont in 2007 and his M.S.~in Environmental Studies, focusing on Conservation Biology, from Antioch University of New England in 2013. He returned to University of Vermont in 2016 to pursue a Ph.D.~in Computer Science as a graduate research fellow in the NSF IGERT Smart Grid program, where his research interests include risk analysis of cascading power failures.
\end{IEEEbiography}

% Got this spacing trick from: https://tex.stackexchange.com/questions/374612/reduce-the-gap-between-bios-in-ieeetran
\vskip -2\baselineskip plus -1fil

\begin{IEEEbiography}[{\includegraphics[width=1in,height=1.25in,clip,keepaspectratio]{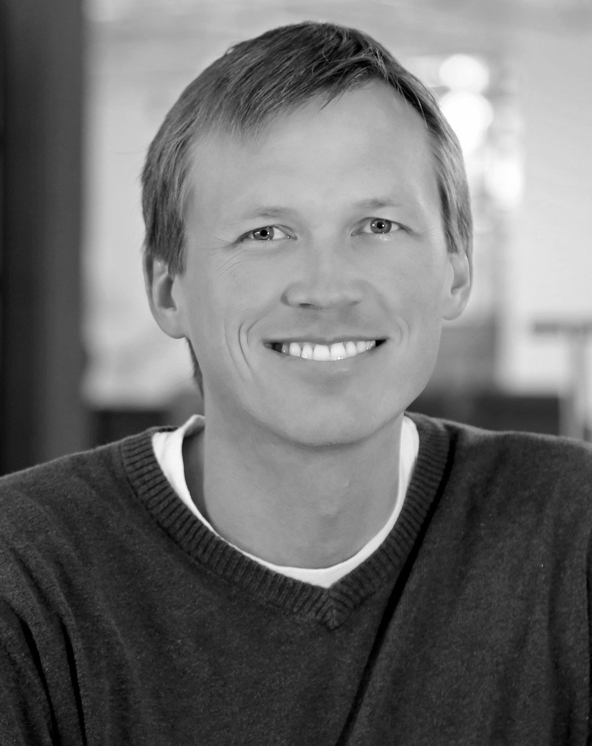}}]{Paul D.H. Hines} (S`96,M`07,SM`14) received the Ph.D.~in Engineering and Public Policy from Carnegie Mellon University in 2007 and M.S.~(2001) and B.S.~(1997) degrees in Electrical Engineering from the University of Washington and Seattle Pacific University, respectively. He is currently Associate Professor and the L. Richard Fisher chair in the Electrical and Biomedical Engineering department, with a secondary appointment in Computer Science, at the University of Vermont. He is also a member of the external faculty of the Santa Fe Institute and a co-founder of Packetized Energy, a cleantech startup. Formerly he worked at the U.S.~National Energy Technology Laboratory, the US Federal Energy Regulatory Commission, Alstom ESCA, and for Black and Veatch. He currently serves as the vice-chair of the IEEE PES Working Group on Cascading Failure.
\end{IEEEbiography}

\vskip -2\baselineskip plus -1fil

%\vfill
%\newpage
%\vfill

%\newpage

\begin{IEEEbiography}[{\includegraphics[width=1in,height=1.25in,clip,keepaspectratio]{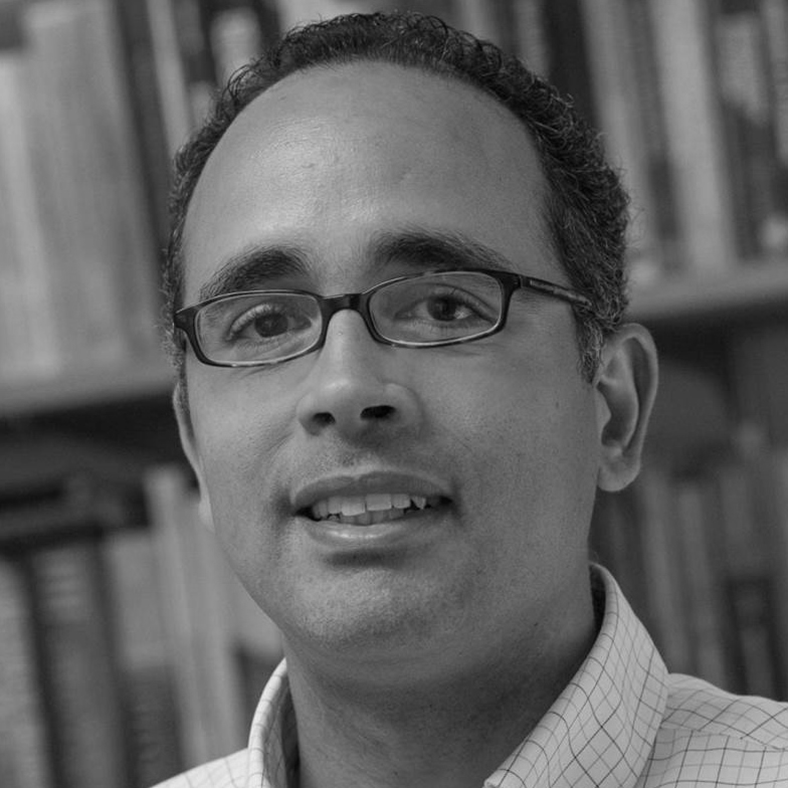}}]{Eric M. Hernandez} received a B.S. degree in Civil Engineering from Universidad Nacional Pedro Henriquez Urena, Santo Domingo, Dominican Republic, in 1999, and M.S. and Ph.D. degrees in Civil Engineering from Northeastern University, in 2004 and 2007, respectively. In 2011, he joined the Department of Civil and Environmental Engineering, at the University of Vermont (UVM) as an Assistant Professor and in 2017 was promoted to Associate Professor. In 2018, he was awarded the inaugural Gregory N. Sweeny Green and Gold Professorship in Civil Engineering at UVM. His current research interests include structural health monitoring, stochastic modeling, inverse problems and Bayesian reliability.
\end{IEEEbiography}

\vskip -2\baselineskip plus -1fil

\begin{IEEEbiography}[{\includegraphics[width=1in,height=1.25in,clip,keepaspectratio]{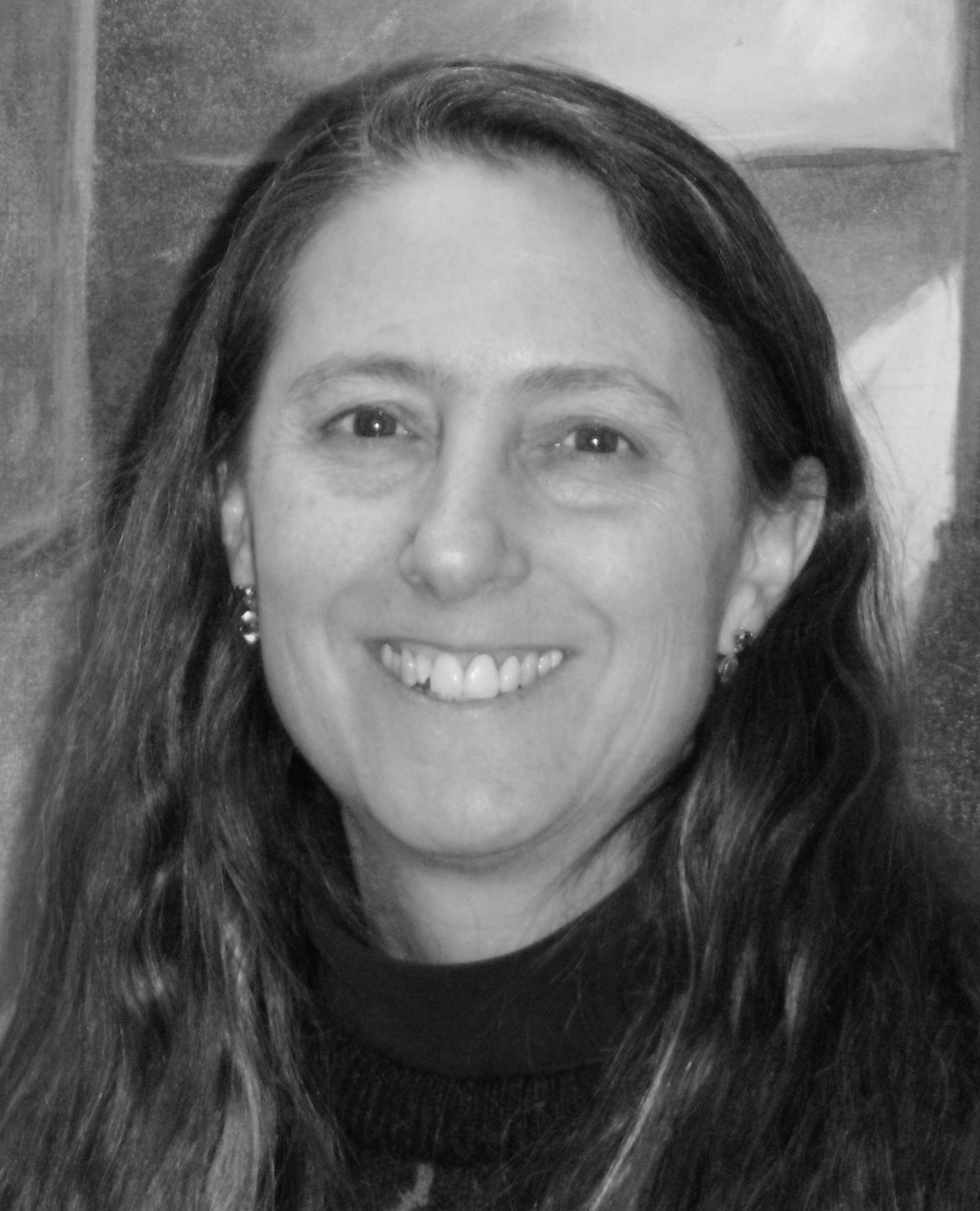}}]{Margaret J. Eppstein} received the B.S. degree in zoology from Michigan State University in 1978 and the M.S. degree in computer science and the Ph.D. degree in civil \& environmental engineering from the University of Vermont (UVM), Burlington, VT, in 1983 and 1997, respectively. She is Research Professor and Professor of Computer Science Emerita at UVM, where she has been on the faculty since 1983. She was founding director of the Vermont Complex Systems Center (2006-2010) and Chair of the UVM Department of Computer Science (2012-2018). Her research interests comprise interesting computational challenges in modeling and analysis of complex systems in a variety of application domains.
\end{IEEEbiography}

\vfill
\vfill


%\vfill


\end{document}